# A Raman Probe of Phonons and Electron-phonon Interactions in the Weyl Semimetal $NbIrTe_4$

*Iraj Abbasian Shojaei [a], Seyyedesadaf Pournia [a], Congcong Le [b,c], Brenden R. Ortiz [d,e], Giriraj Jnawali [a], Fu-Chun Zhang [b], Stephen D. Wilson [d,e], Howard E. Jackson [a], Leigh M. Smith* [a],*

[a] *Department of Physics, University of Cincinnati, Cincinnati, OH, USA*
[b] *Kavli Institute of Theoretical Sciences, University of Chinese Academy of Sciences, Beijing 100190, China*
[c] *Max Planck Institute for Chemical Physics of Solids, 01187 Dresden, Germany*
[d] *Materials Department, University of California Santa Barbara, Santa Barbara CA 93106*
[e] *California Nanosystems Institute, University of California Santa Barbara, Santa Barbara CA 93106*

*email: leigh.smith@uc.edu

## Abstract

There is tremendous interest in measuring the strong electron-phonon interactions seen in topological Weyl semetals. The semimetal $NbIrTe_4$ has been proposed to be a Type-II Weyl semimetal with 8 pairs of opposite Chirality Weyl nodes which are very close to the Fermi energy. We show using polarized angular-resolved micro-Raman scattering at two excitation energies that we can extract the phonon mode dependence of the Raman tensor elements from the shape of the scattering efficiency versus angle. This van der Waals semimetal with broken inversion symmetry and 24 atoms per unit cell has 69 possible phonon modes of which we measure 19 modes with frequencies and symmetries consistent with Density Functional Theory calculations. We show that these tensor elements vary substantially in a small energy range which reflects a strong variation of the electron-phonon coupling for these modes.

## Introduction

Three dimensional topological semimetals have attracted intense interest in recent years.[1]–[5] These materials are enabled by broken spatial or time inversion symmetries, and the large spin orbit interactions which collapses the conduction and valence bands within a limited section of the Brillouin zone where the conduction and valence bands cross and invert. Dirac, Weyl, nodal-line, and nested fermion semimetals have all been predicted theoretically and observed experimentally.[6]–[11] Weyl semimetals are particularly interesting since pairs of Weyl nodes with opposite chirality appear as monopole sources and sinks of Berry curvature.[12], [13]

This results in unusual transport properties such as the anomalous Hall effect, extremely large and non-saturating magneto-resistance, the chiral anomaly and negative longitudinal magneto-resistance.[14]–[19] In addition, extremely large nonlinear optical effects such as sum and difference frequency generation, nonlinear DC shift currents (electromagnetic rectification), and photoinduced anomalous Hall effect have been predicted and observed.[20]–[25] The Weyl semimetals are classified in two flavors: Type-I, which approximately maintain Lorentz invariance, and Type-II where the linear crossings are tilted as to result in electron and hole pockets in the Fermi surface at the Weyl point.[26]–[28]

Recently, both theoretical and experimental evidence demonstrate that the electron-phonon interactions in Weyl semimetals can provide insights into the electronic structure of these unusual materials. For example, calculations show that one can use magnetic field effects on phonons in Weyl semimetals in both enantiomorphic[29] and mirror-symmetric[30] materials as a direct result of the chiral anomaly. Phonon-mediated changes in hydrodynamic flow in WTe2 have also very recently been measured.[31] Changes in the phonon dispersion, optical reflectivity and Raman scattering are expected.[29], [30], [32], [33] Evidence for the strong coupling between phonons and Weyl Fermions is seen by Fano resonances observed in temperature tuned TaAs (IR reflectivity)[34], temperature-dependent Raman measurements in NbAs, TaAs and WP2,[35], [36] and also in Resonant Raman Scattering in TaP[37]. The chiral anomaly in NbAs was seen in phonon measurements using magnetic field-dependent IR reflectivity measurements[38]. In this paper we use polarized angular-resolved Raman scattering in a Type II Weyl semimetal to show that one can extract the Raman tensor elements for each phonon mode from the *shape* of the scattering efficiency versus angle. Thus the electron-phonon coupling for individual phonon modes can be measured for different excitation energies.

Weyl semimetals are only possible for materials which break either spatial inversion symmetry or time inversion symmetry, or both.[3] The two-dimensional $NbIrTe_4$ material is a ternary alloy which is analogous to the known Type-II Weyl semimetal $WTe_2$, so it is expected to break inversion symmetry. Several *ab initio* calculations including spin-orbit coupling indicate that $NbIrTe_4$ is a type-II Weyl semimetal with 16 Weyl points which contains 8 pairs of opposite chirality located at energies very close to the Fermi energy.[39] With 24 atoms per unit cell, 69 phonon modes are possible in this structure.[40] While several magneto-transport measurements in late 2019 have shown non saturating magneto-resistance at low temperature, and quantum oscillations[41], [42] which are consistent with a complex Fermi surface, there are no Angle-Resolved Photoemission Spectroscopy (ARPES) measurements to investigate the expected surface Fermi arcs or band crossings expected in this material.

In the first section, we describe angle-resolved polarized micro-Raman scattering measurements on a nanoflake of $NbIrTe_4$ to investigate the symmetry and frequency of the Stokes scattered Raman modes. We show that the observed modes and symmetries are consistent with *ab-initio* Density Functional Theoretical (DFT) calculations and extract the normal modes of atomic vibrations for these modes.

Then we consider whether there is evidence of variable electron phonon interactions for different laser excitations. We compare in detail the symmetries and intensities of the observed Raman active modes in $NbIrTe_4$, which are excited at 633 nm (red) and at 514 nm (green) wavelengths. We observe enhanced intensity for most Raman modes for 633 nm excitation, and also a notable variation between the two excitation wavelengths for different rotational symmetries of the modes which identify changes in the electron-phonon interaction. These changes with excitation wavelength vary substantially from mode to mode. We conclude with a discussion which shows that we can extract detailed information about each of the Raman tensor elements (electron-phonon coupling constants) for 13 phonon modes ranging from 45 to 260 $cm^{-1}$ (5 to 32 meV).

**Sample Morphology and Experimental Setup**

A single crystal ternary $NbIrTe_4$ alloy was grown by the flux method (see Methods). The orthorhombic primitive cell of this structure (Fig. 1a) includes 24 atoms with four formula units.[43] Chains of Nb and Ir atoms alternatively follow a zigzag pattern along the 'a' axis, and hybridize with each other long the 'b' axis to form a conducting plane. Te atoms bond on top and bottom of the 'ab' planes to form layers which are van der Waals bonded along the 'c' direction.[43]

A transmission electron microscopy (TEM) image and electron diffraction pattern of $NbIrTe_4$ sample taken along the 'c' direction is illustrated at Fig. 1b. The TEM diffraction pattern is taken with the electron beam aligned with the central part of a $NbIrTe_4$ exfoliated flake dispersed onto a holey carbon grid. The TEM image exhibits the uniformity of the exfoliated flake, while the diffraction pattern confirms the symmetry of the orthorhombic structure of this material with the a and b directions of the diffraction pattern corresponding to the 'a' and 'b' axes of the flake.

X-ray diffraction measurements (XRD) from a $NbIrTe_4$ single crystal grown in the same way determined lattice constants of 3.79(03), 12.52(07) and 13.14(35) Angstroms for the 'a', 'b' and 'c' lattice constants, respectively. These lattice constants are very close to the calculated values in the literature[43] and are used as inputs to our DFT calculation (see Methods) in order to obtain appropriate interatomic potentials for this crystal. Details of the DFT calculations are in the supplementary document (S1).

The nanoflakes were exfoliated from a $NbIrTe_4$ single crystal by using NITTO blue tape, and then mechanically transferred on a $Si/SiO_2$ substrate for angle-resolved polarized micro-Raman spectroscopy (see optical image in Fig. 1c).

Figure 1d illustrates a schematic diagram of our polarized micro-Raman experiment system (see Methods). A linearly polarized 632.9 nm or 514.5 nm laser is focused to a ~1.5 μm spot onto the nanoflake with the optic axis aligned parallel to the 'c' axis of the nanoflake. The substrate with the nanoflake is rotated keeping the laser spot fixed so that the polarization is at an angle θ relative to the 'a' axis (Fig. 1c). The double-sided black arrow in the image shows the polarization direction of the incident light relative to the 'a' and 'b' axes of the nanoflake marked by white arrows. A polarization analyzer is set so that the polarization of the scattered light ($e_s$) is either parallel ($e_i \parallel e_s$) or perpendicular ($e_i \perp e_s$) to the incident laser polarization ($e_i$). Data for both configurations is collected with a 10 minute exposure time with the sample rotated through 360 degrees with 10-degree increments.

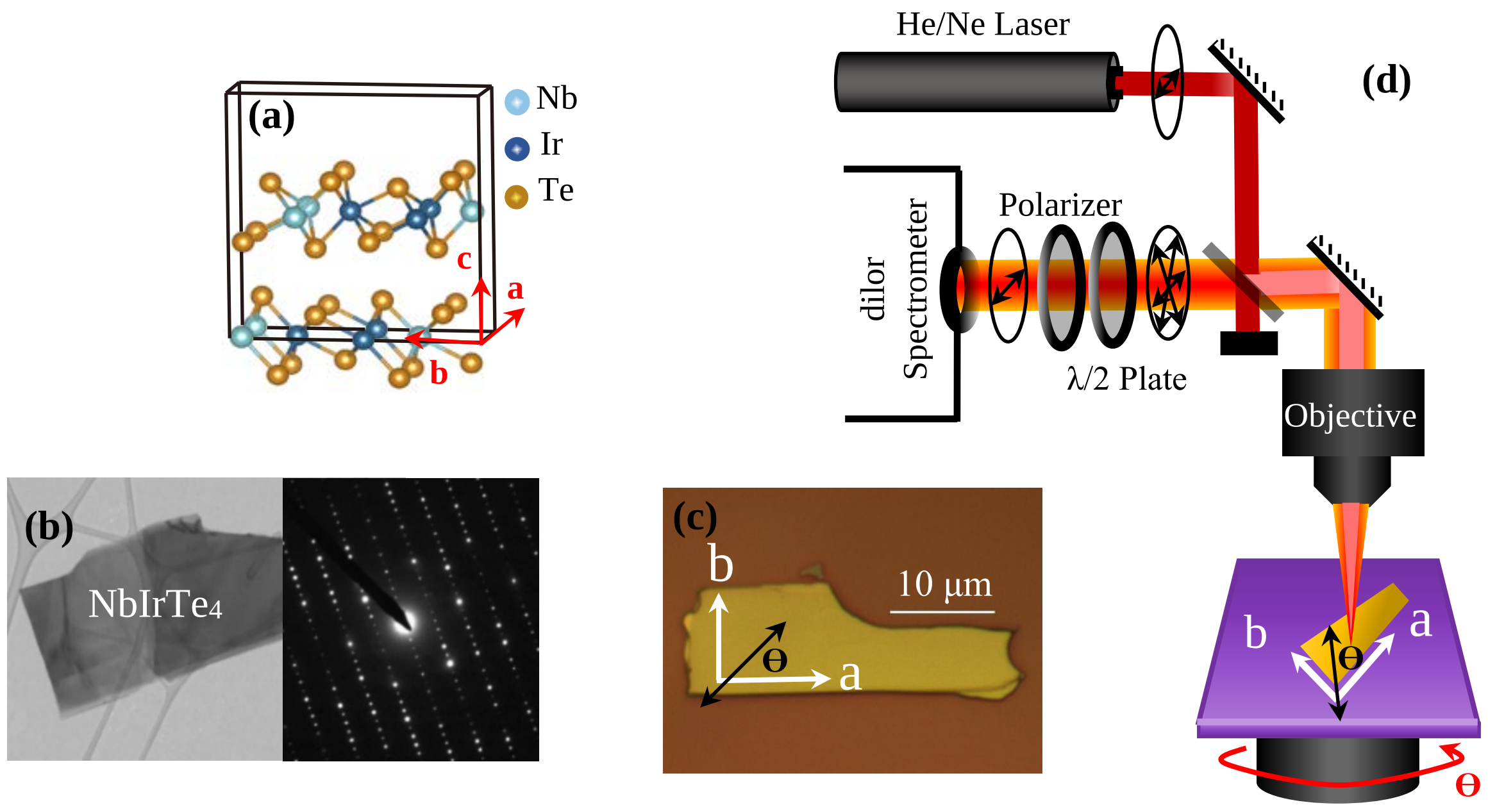

Figure 1: a) Schematic side perspective of crystal structure of $NbIrTe_4$ b) TEM image and TEM diffraction pattern of $NbIrTe_4$ along c-axis c) The optical image of the $NbIrTe_4$ sample including ‘a’ and ‘b’ crystal line axes and the polarization direction of the incident light d) The schematic diagram of polarized Raman measurement with rotational stage.

## Experimental and Theoretical Results

The orthorhombic primitive cell of $NbIrTe_4$ structure belongs to the space group $Pmn2_1$ or point group $C_{2v}$ including 4 irreducible representations, $A_{1,2}$ and $B_{1,2}$. From the character table, the $NbIrTe_4$ unit cell includes 24 atoms so that 72 phonon modes are possible with 69 Raman-active modes: 23 $A_1$ + 12 $A_2$ + 11 $B_1$ + 23 $B_2$. Detecting modes with different irreducible representation can be achieved by selecting the laser excitation direction along with the polarization configuration of the experiment.

The incoming and backscattered beams are along the ‘c’ axis of $NbIrTe_4$ lattice (see Fig. 1d), and perpendicular to the nanoflake. We align the ‘x’, ‘y’ and ‘z’ directions of lab coordinate system with the ‘a’, ‘b’ and ‘c’ crystallographic axes of the structure, respectively. The angle θ is thus between the polarization direction of the incident beam and the ‘x’ axis. The Raman experiments are taken either in the parallel configuration ($e_i \parallel e_s$) or in the perpendicular configuration ($e_i \perp e_s$.)

We recall that the scattered intensity in Raman experiments is proportional to the excitation frequency and polarization direction of the incident and scattered light[44]:

$$I(\theta, \varphi, \omega_s) = A\,\omega_s^4\,|\mathbf{e_i}.\mathbf{R}.\mathbf{e_s}|^2 \qquad (1)$$

$$R \sim \sum_{\alpha,\beta} \frac{< f|H_{eo}|\beta >< \beta|H_{ep}|\alpha >< \alpha|H_{eo}|i >}{(E_L - (E_\beta - E_i) - \hbar\omega_v - i\gamma)(E_L - (E_\alpha - E_i) - i\gamma)}$$

where A provides the scaling with incident laser power, $\omega_i$ and $\omega_s$ are the frequencies of the incident and scattered photons, **R** is the Raman tensor reflecting selection rules and containing the electron-phonon coupling parameters, $\mathbf{e_i}$ and $\mathbf{e_s}$ are the incident and scattered polarization vectors. $E_L$ is the incident laser energy, $E_\alpha$ and $E_\beta$ are the energies of the intermediate electronic states $|\alpha>$ and $|\beta>$, $\omega_v$ is the phonon frequency, $E_i$ is the energy of the initial electronic state. $H_{eo}$ and $H_{ep}$ are the electron-photon and electron-phonon operators. Thus the denominator describes resonance conditions for the incoming and outgoing photons.[44] After removing the $\omega_s^4$ dependence, the normalized scattered intensities at two different excitation wavelengths provide a view into the Raman tensor dependence on crystal orientation and excitation energy and geometry for each mode and thus connect that to the electron-phonon interactions in this material. The Raman tensors corresponding to the $A_1$, $A_2$, $B_1$ and $B_2$ irreducible representations are given by [45], [46]

$$R_{A_1} = \begin{pmatrix} d & 0 & 0 \\ 0 & f & 0 \\ 0 & 0 & g \end{pmatrix}\; R_{A_2} = \begin{pmatrix} 0 & h & 0 \\ h & 0 & 0 \\ 0 & 0 & 0 \end{pmatrix}\; R_{B_1} = \begin{pmatrix} 0 & 0 & 0 \\ 0 & 0 & k \\ 0 & k & 0 \end{pmatrix}\; R_{B_2} = \begin{pmatrix} 0 & 0 & l \\ 0 & 0 & 0 \\ l & 0 & 0 \end{pmatrix} \qquad (2)$$

where d, f, g, h, k and l are complex tensor elements of **R** which depend on the derivatives of the complex dielectric response and are connected to the electron-phonon coupling between electronic and lattice states.[47] Each complex element includes a phase term in the form of the $x = |x|e^{i\varphi_x}$.

According to Porto's notation, we see from equation (2) that the $Z(XX)\bar{Z}$ and $Z(YY)\bar{Z}$ measurements ($\theta = 0^o$ and $\theta = 90^o$ when $e_i \parallel e_s$) reveal $A_1$ modes and, $Z(XY)\bar{Z}$ and $Z(YX)\bar{Z}$ measurements ($\theta = 0^o$ and $\theta = 90^o$ when $e_i \perp e_s$) reveal $A_2$ modes. Also, because the incident and scattered light propagates along 'c' (z) axis direction (the electric fields are parallel to the xy plane for any θ), Raman experiments do not probe the $B_1$ or $B_2$ phonon modes.

For the incident laser polarization at an angle of θ relative to the a axis, the elements of $\mathbf{e_i}$ and $\mathbf{e_s}$ are (cos(θ), sin(θ), 0) for the $e_i \parallel e_s$ a measurement, while for the $e_i \perp e_s$ measurement $\mathbf{e_s}$ has (-sin(θ), cos(θ), 0) elements. Thus, at a fixed excitation frequency, by using $\mathbf{e_i}$ and $\mathbf{e_s}$ in equation (1), the intensity of Raman modes as a function of rotation angle will take the form:

$$I_{A_1}^{\parallel} = |d|^2\cos^4(\theta) + |f|^2\sin^4(\theta) + 2|d||f|\cos^2(\theta)\sin^2(\theta)\cos(\varphi_{df}) \qquad (3)$$

$$I_{A_1}^{\perp} = \cos^2(\theta)\sin^2(\theta)[|d|^2 + |f|^2 - 2|d||f|\cos(\varphi_{df})] \qquad (4)$$

$$I_{A_2}^{\parallel} = |h|^2\sin^2(2\theta) \qquad (5)$$

$$I_{A_2}^{\perp} = |h|^2\cos^2(2\theta) \qquad (6)$$

$$I_{B_1}^{\parallel} = I_{B_1}^{\perp} = I_{B_2}^{\parallel} = I_{B_2}^{\perp} = 0 \qquad (7)$$

where $\varphi_{df} = \varphi_d - \varphi_f$ is the relative complex phase factor between the d and f tensor elements. From this result we see that the angular dependence of the $A_1$ intensity when measured in the parallel configuration depends strongly on the d and f elements, which reflect the electron-phonon coupling. For all other measurements, the angular dependence only depends on geometry. The $|d|/|f|$ ratio and $\varphi_{df}$, thus have an important role in the angular behavior of the $A_1$ Raman modes which is discussed in supplementary documents (S2, S3). [47]

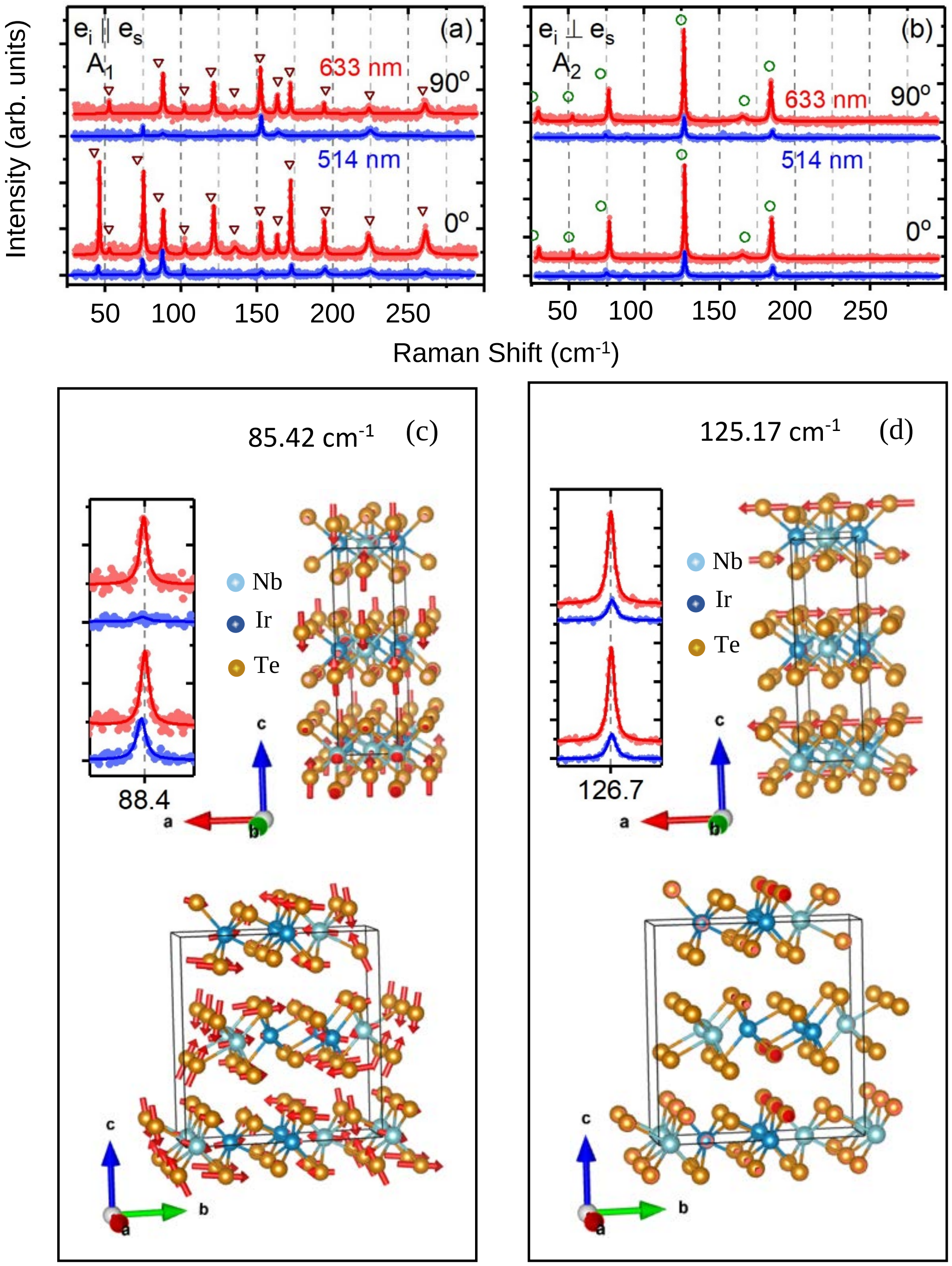

Figure 2: Polarized Raman spectra of $NbIrTe_4$ nano-flake as a function of rotation angle. a, b) Raman spectra for two selected angles ($0^o$, $90^o$) between polarization direction of the incident beam and a-axis of the sample, respectively, for parallel and cross configurations. Red and blue lines are Lorentz fittings of the Raman spectra measured with 633 nm and 514 nm laser excitation, respectively. Dark red triangles and green circles illustrate the calculated Raman frequencies of the detected modes by Density Functional Theory belonging to the $A_1$ and $A_2$ irreducible representations, respectively. c, d) Simulated normal modes of given frequencies calculated by the Density Function Theory at two perspectives for a one selected $A_1$ mode at 85.42 $cm^{-1}$ and one selected $A_2$ mode at 125.17 $cm^{-1}$, respectively. The direction of arrows shows the direction of the atoms vibration and magnitude of arrows represent relative amplitude of atoms vibration.

As noted previously, (equations (3) to (6)), one can distinguish that for $Z(XX)\bar{Z}$ and $Z(YY)\bar{Z}$ measurements only $A_1$ modes can be seen while for $Z(XY)\bar{Z}$ and $Z(YX)\bar{Z}$ measurements only the $A_2$ modes are observed. Fig. 2a and 2b show Raman spectra for ($Z(XX)\bar{Z}$, $Z(YY)\bar{Z}$) and ($Z(XY)\bar{Z}$, $Z(YX)\bar{Z}$) respectively, with a Lorentz fit to all spectral lines observed for 633 nm (red curve) and 514 nm (blue curve) excitations. 13 $A_1$ and 6 $A_2$ Raman modes are observed. These measurements have been calibrated utilizing a helium gas discharge near 514 and 633 nm which provide a spectral resolution and linewidth of around 1 $cm^{-1}$ (see S4). The linewidths of the detected Raman modes are around 1 $cm^{-1}$ (limited by our instrument resolution) except 2 modes at 224.1 and 261.2 $cm^{-1}$ which show a 3.5 linewidth. This confirms the high crystal purity of our samples. Comparison to DFT calculations for both $A_1$ and $A_2$ modes is exhibited by dark red triangles and green circles, respectively. Except for 3 $A_1$ modes (42.84 $cm^{-1}$, 52.52 $cm^{-1}$, 71.35 $cm^{-1}$) and 2 $A_2$ modes ( 26.55 $cm^{-1}$, 71.59 $cm^{-1}$) which are within 7% of the DFT frequencies, the remaining 14 detected mode frequencies agree within 2% or less of the DFT frequencies. The exact values of our measurements for $A_1$ and $A_2$ Raman frequencies and DFT calculations are listed in Table 1. In this table all 35 $A_{1,2}$ modes predicted by DFT calculations are listed. The predicted values of the $B_{1,2}$ modes can be found in the in the supplementary documents (S5). The details of the DFT calculations which provide both the frequency and symmetry of active Raman modes are provided in the supplementary documents (S1).

Fig. 2c and 2d show spectra and fits to a representative $A_1$ and $A_2$ mode at both red (633 nm) and green (514 nm) excitations. The frequencies extracted from these fits are 88.4 $cm^{-1}$ for the $A_1$ mode and 126.7 $cm^{-1}$ for the $A_2$ mode. For comparison, the frequency of the DFT calculated mode is shown as symbols: 85.42 $cm^{-1}$ for the $A_1$ mode (3% error) and 125.17 $cm^{-1}$ for the $A_2$ mode (1% error). The atomic vibrations for these two modes are shown with two different perspectives, namely viewed along either the 'a' or 'b' axis. Note that the motion of atoms for the $A_1$ modes is restricted to the 'cb' crystal plane, while the atoms for the $A_2$ modes can only vibrate parallel to the 'a' axis of the crystal. This is true for all modes with these symmetries. For the $A_1$ mode at 85.42 $cm^{-1}$ all Nb, Ir and Te atoms are seen to move, while for the $A_2$ mode at 125.17 $cm^{-1}$ only some of the tellurium atoms move. The details about the atomic vibrations for other modes are in the supplementary documents (S6).

| Mode Type in $Pmn2_1$ | DFT Calculation ($cm^{-1}$) | $e_i \parallel e_s$ 633nm ($cm^{-1}$) | $e_i \parallel e_s$ 514nm ($cm^{-1}$) | $e_i \perp e_s$ 633nm ($cm^{-1}$) | $e_i \perp e_s$ 514nm ($cm^{-1}$) | $e_i \parallel e_s$ $I_{633}/I_{514}$ | \|d\|/\|f\| 633nm | \|d\|/\|f\| 514nm | $\varphi_{df}$ (°) 633nm | $\varphi_{df}$ (°) 514nm |
|---|---|---|---|---|---|---|---|---|---|---|
| $A_1$ | 42.84 | 46.4 | 45.6 | 46.2 | 45.9 | 8.3 | 2.5 | 1.2 | 94 | 64 |
| $A_1$ | 52.52 | - | - | 56.3 | - | - | 1 | - | 90 | - |
| $A_1$ | 71.35 | 75.4 | 75.2 | 75.4 | 75.6 | 4.6 | 3.1 | 1.5 | 105 | 96 |
| $A_1$ | 81.45 | - | - | - | - | - | - | - | - | - |
| $A_1$ | 85.42 | 88.4 | 88.0 | 88.4 | 88.0 | 1.7 | 1 | 2.1 | 90 | 84 |
| $A_1$ | 102.16 | 102.3 | 102.1 | 102.3 | - | 1.3 | 1 | 1.8 | 90 | 12 |
| $A_1$ | 107.51 | - | - | - | - | - | - | - | - | - |
| $A_1$ | 119.86 | 121.7 | - | 121.7 | - | $\infty$ | 1.2 | - | 47 | - |
| $A_1$ | 134.01 | - | - | - | - | - | - | - | - | - |
| $A_1$ | 135.31 | 135.9 | - | 135.9 | - | $\infty$ | 1.4 | - | 130 | - |
| $A_1$ | 139.86 | - | - | - | - | - | - | - | - | - |
| $A_1$ | 152.11 | 152.5 | 153.2 | 152.5 | 153.3 | 2.5 | 0.8 | 0.6 | 120 | 112 |
| $A_1$ | 162.47 | - | - | - | - | - | - | - | - | - |
| $A_1$ | 163.97 | 164.0 | 164.0 | - | - | 3 | 1 | 0.8 | 20 | 47 |
| $A_1$ | 171.64 | 172.4 | 173.4 | 172.4 | - | 7.1 | 1.5 | 2.3 | 50 | 72 |
| $A_1$ | 187.48 | - | - | - | - | - | - | - | - | - |
| $A_1$ | 192.10 | - | - | - | - | - | - | - | - | - |
| $A_1$ | 195.31 | 194.4 | 195.0 | 194.4 | - | 3.7 | 1.6 | 1.9 | 54 | 44 |
| $A_1$ | 220.88 | - | - | - | - | - | - | - | - | - |
| $A_1$ | 224.53 | 224.1 | 225.2 | 224.1 | - | 2.5 | 1.4 | 1 | 51 | 32 |
| $A_1$ | 259.45 | 261.2 | 260.7 | 261.2 | - | 6.5 | 1.2 | 1.2 | 94 | 10 |
| $A_1$ | 266.05 | - | - | - | - | - | - | - | - | - |
| $A_2$ | 26.55 | - | - | 30.3 | - | - | | | | |
| $A_2$ | 50.23 | - | - | 52.6 | - | - | | | | |
| $A_2$ | 56.60 | - | - | - | - | - | | | | |
| $A_2$ | 71.59 | 77.0 | - | 77.0 | - | - | | | | |
| $A_2$ | 99.57 | - | - | - | - | - | | | | |
| $A_2$ | 103.68 | - | - | - | - | - | | | | |
| $A_2$ | 124.54 | - | - | - | - | - | | | | |
| $A_2$ | 125.17 | 126.7 | 126.8 | 126.7 | 126.8 | 2.5 | | | | |
| $A_2$ | 159.87 | - | - | - | - | - | | | | |
| $A_2$ | 166.94 | - | - | 165.0 | - | - | | | | |
| $A_2$ | 182.00 | - | - | - | - | - | | | | |
| $A_2$ | 183.83 | 184.5 | 185.3 | 184.35 | 185.3 | 2.8 | | | | |

Table 1: Calculated $A_1$ and $A_2$ Raman modes of the $NbIrTe_4$ nano-flake by using Density Functional Theory (DFT) and experimentally extracted values in parallel and cross configurations using 633 nm and 514 nm excitation. The last four columns express the parameters d/f and $\phi_{df}$ determined from the angular dependence for each experimentally detected $A_1$ modes.

Comparing the Raman spectra in Fig. 2 obtained for green (514 nm) or red (633 nm) laser excitation, one can immediately see that generally the scattering efficiencies for most modes increase strongly for red excitation. From equation (1) the scattering efficiency usually is dominated by the $\omega^4$ term which thus would increase the intensities for green excitation. In order to see clearly the change in the matrix elements, the spectra in this paper are normalized to the measured integrated intensity of the well-known 520 $cm^{-1}$ silicon (optical phonon) Raman line. This removes both the $\omega^4$ term and any dependence of the measurement from the Raman instrument and detector. By comparison with reflectivity measurements in the closely related materials $TaIrTe_4$ [48] and $WTe_2$ [49] we do not expect any large change in the dielectric response (real or imaginary parts) with laser excitation energy which might also impact the Raman measurements. The penetration depth is also substantially less than the nanoflake thickness. The normalized intensity ratios ($I_{633}/I_{514}$) for each mode are recorded in Table 1. While two modes at 85.42$cm^{-1}$ and 102.16 $cm^{-1}$ do not show significant change, the two modes at 119.86 $cm^{-1}$ and 135.31 $cm^{-1}$ are not detected for 514 nm excitation at all. There are no modes which show a weaker response for red laser excitation, and most show an enhancement of the matrix element by a factor of 2 to 8. As we shall now discuss, there are also strong changes in the angular dependence of $A_1$ modes measured for $e_i \parallel e_s$ for different excitation energies.

## The Raman Tensor, Raman Intensities and variation of electron phonon interaction

As captured by equations 1 -7, the intensity of the Raman scattering for different modes depends on both the values of |d|, |f|, and |h| as well as the geometry as reflected in the angle $\theta$. We will see that presentation of the data in polar plots reflects this variation and note that, as well, the response can be significantly different for the two different excitation wavelengths.

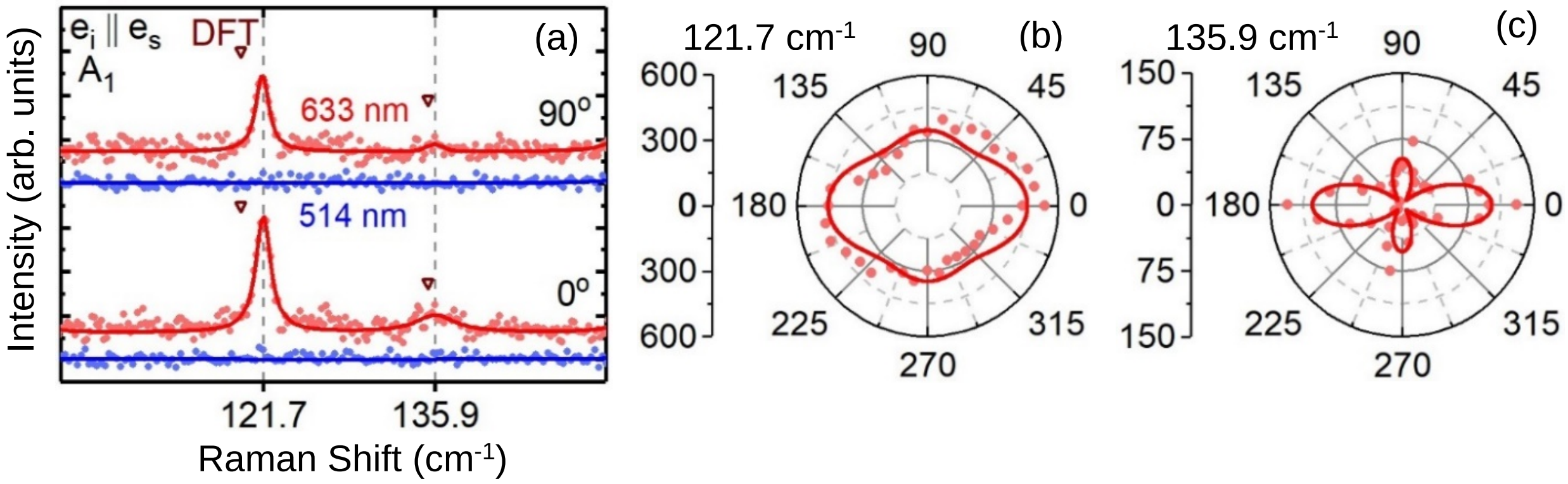

Figure 3: a) Raman spectra for two peaks at angles of 0º, 90º for the parallel polarization configuration with 633 nm excitation (red lines) and 514 nm (blue lines) excitation. Dark red triangles mark the calculated modes utilizing Density Function Theory. b, c) Polar plots of the Raman modes at 121.7 $cm^{-1}$ and 135.9 $cm^{-1}$ respectively using 633 nm excitation.

Fig. 3a illustrates a segment of our $Z(XX)\bar{Z}$ and $Z(YY)\bar{Z}$ Raman measurements where two 121.7 and 135.9 $cm^{-1}$ modes are only observed with 633 nm excitation (red dots and line); no response of these modes is observed for 514 nm excitation measurements (blue dots and line). Thus, these two modes clearly demonstrate a strong dependency on excitation energy suggesting that the coupling between electronic states and lattice states is very weak for 514 nm excitation. Fitting spectra of our angle-resolved measurements with the $e_i$ || $e_s$ configuration for these two modes is obtained by fixing the frequency and varying only the linewidth and intensity for all fits. The results of the intensity versus angle are fit to equation (3), as shown with solid lines in Fig. 3b and 3c. As detailed in the supplementary (S2) , these plots allow us to extract the relative value of '|d|' and '|f|' for each mode. For instance, for the mode at 121.7 $cm^{-1}$, |d| and |f| are nearly equal. For the mode at 135.9 $cm^{-1}$ |d| is about 40% larger than |f|.

The other observed 10 $A_1$ modes can classify into two categories. In one category, regardless the effect of $\varphi_{df}$ on rotational symmetry of the modes (S3), the |d|/|f| ratio stays either larger than or smaller than 1 as the excitation energy is changed. Fig. 4 shows polar plots of 6 modes at two 633 nm (red curves) and 514 nm (blue curves) excitations for a $e_i$ || $e_s$ configuration. This figure displays 5 modes at 46.4, 75.4, 172.4,194.4 and 261.2 $cm^{-1}$ where the maximum for both excitation wavelengths is along 'x' direction with the ratio of |d|/|f| larger than 1 (S2) for both excitations. We see that the maximum of plot at 152.5 $cm^{-1}$ is along 'y' direction which means |d|/|f| is smaller than 1 for both excitations.

For each of these 10 modes, we note that the integrated intensity with 633 nm excitation is *larger* than for the comparable 514 nm measurements. The blue numbers within each plot show the factor by which the of 514 nm measurements are multiplied to have equal intensity with 633 nm measurements. As noted in Table 1, this factor is different for each mode, with the largest value of 8.3 for the mode at 46.4 $cm^{-1}$ and the smallest value of 1.3 for 102.2 $cm^{-1}$.

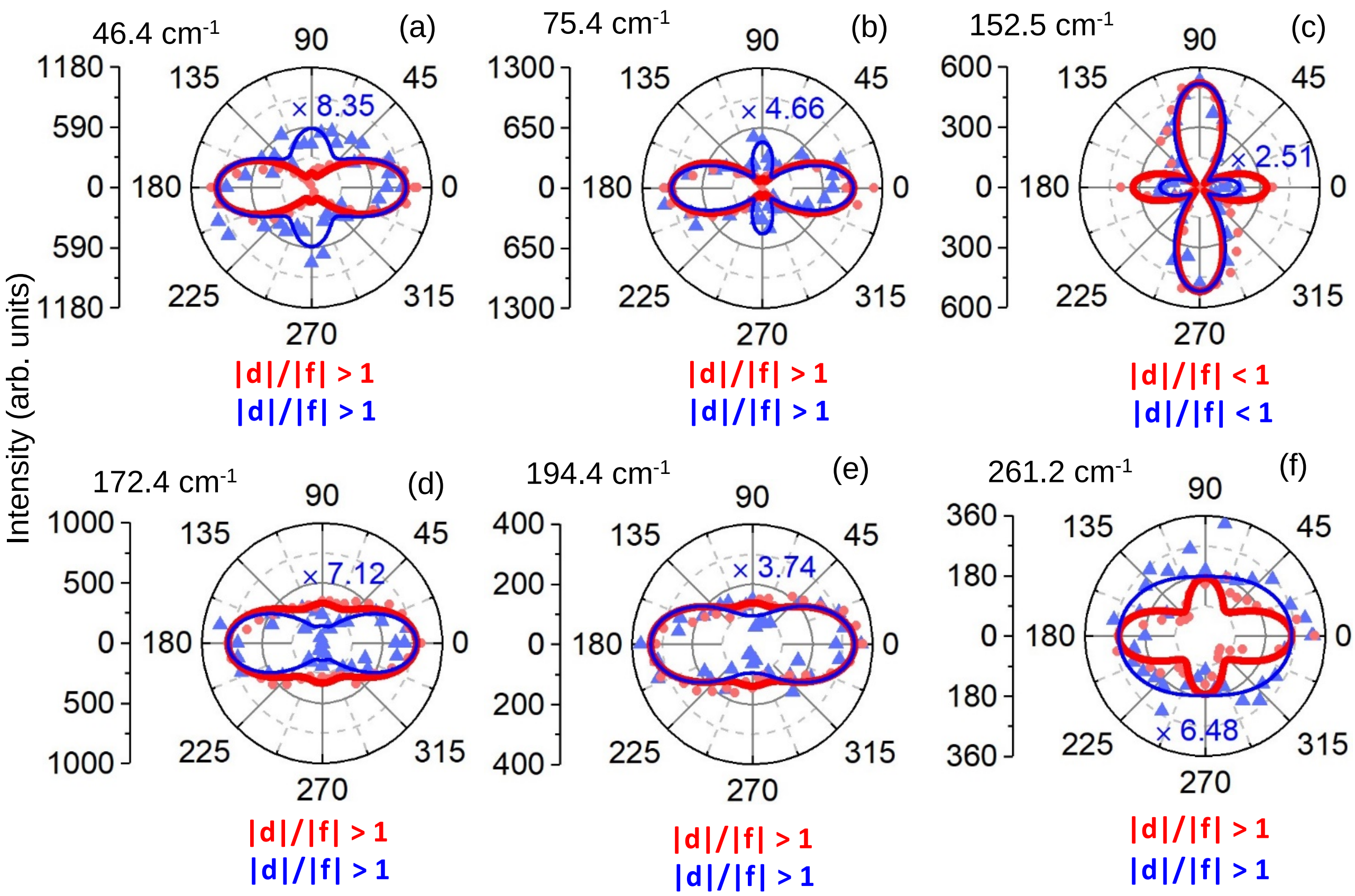

Figure 4: Intensity polar plots (a – f) of six observed Raman modes of $NbIrTe_4$ with a |d|/|f| ratio of larger or smaller than 1 for both excitation wavelengths. In (c) the ratio of |d|/|f| is smaller than 1 and all others are larger than 1. Red (blue) curves illustrate theoretical fitting of measurements for 633 nm (514 nm) excitation. To facilitate comparison, a magnification is applied for each Raman mode measured at 514 nm excitation (blue number within plot).

In the second category, which include 4 (different) modes, the |d|/|f| ratio *equals* 1 for one excitation and is distinctly different from one for another excitation. Fig. 5 shows polar plots of 4 modes at both excitation wavelengths for $e_i \parallel e_s$ configuration. One can see that the |d|/|f| ratio equals 1 for the 633 nm measurements for modes at 88.4, 102.3, and 164 $cm^{-1}$ modes; in contrast, for 514 nm excitation |d|/|f| is larger than 1 at 88.4 (|d|/|f| = 2.1) and 102.3 $cm^{-1}$ (|d|/|f| = 1.8) and is smaller than 1 at 164 $cm^{-1}$ (|d|/|f| = 0.8) for 514 nm excitation. For mode at 224.1, |d|/|f| ratio equals 1 for the 514 nm measurements while for 633 excitation |d|/|f| equals 1.5. Clearly coupling between electron and lattice states strongly depends on excitation energy at these modes.

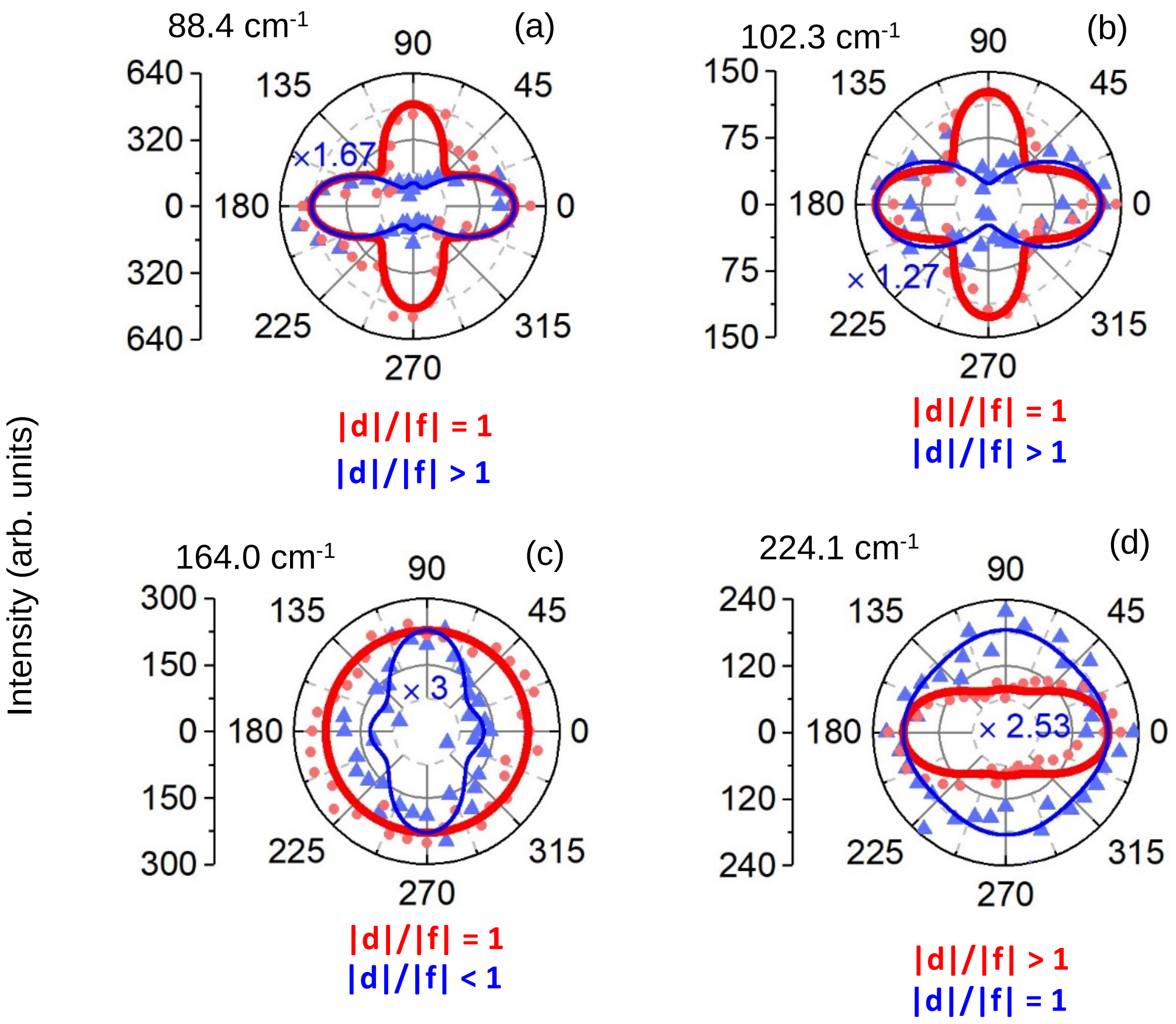

Figure 5: Intensity polar plot of four observed Raman modes of $NbIrTe_4$ with |d|/|f| ratio equals 1 for one of the excitation energies. Plots (a-c) show modes where just for measurement of the 633 nm excitation the ration of |d|/|f| equals 1, and plot (d) shows the mode where just for measurement of the 514 nm excitation the ration of |d|/|f| equals 1. Red (blue) curves illustrate theoretical fitting of measurement by 633 nm (514 nm) laser beam. The factor of magnification for each Raman mode measured by 514 nm laser is declared by blue number at each plot.

## Discussion

In the experiments described above, we have explored Raman scattering from 13 $A_1$ modes and 6 $A_2$ modes which are excited using both green (514 nm) and red (633 nm) laser excitation. We have shown that we can extract |d|/|f| and $\phi_{df}$ for every $A_1$ mode by fitting the *shape* of the angular dependence of the Raman spectra for ($e_i||e_s$) using equation (3). It is straightforward to show that by combining this data with the intensity ratios, $I_{633}/I_{514}$, one can extract, directly the excitation energy dependence of |d| and |f| directly (S7). Specifically, for $A_1$ modes which exhibit a maximum intensity for $\theta = 0^o$, we find:

$$\frac{|f_{633}|}{|f_{514}|} = \sqrt{\frac{I_{633}}{I_{514}} \frac{(|d_{514}|/|f_{514}|)}{(|d_{633}|/|f_{633}|)}} \qquad (8)$$

If the maximum intensity for the mode is at $\theta = 90^o$ then:

$$\frac{|f_{633}|}{|f_{514}|} = \sqrt{\frac{I_{633}}{I_{514}}} \qquad (9)$$

From this we can extract the ratio for the d Raman tensor element:

$$\frac{|d_{633}|}{|d_{514}|} = \left(\frac{|f_{633}|}{|f_{514}|}\right)\frac{(|d_{633}|/|f_{633}|)}{(|d_{514}|/|f_{514}|)} \qquad (10)$$

For the $A_2$ modes where we can measure the intensity ratios $I_{633}/I_{514}$ which is directly related to the h Raman tensor element:

$$\frac{|h_{633}|}{|h_{514}|} = \sqrt{\frac{I_{633}}{I_{514}}} \qquad (11)$$

In Table 2 we show all of the results from these experiments.

| Symmetry | f ($cm^{-1}$) | $I_{633}/I_{514}$ | $\lvert d_{633}\rvert/\lvert d_{514}\rvert$ | $\lvert f_{633}\rvert/\lvert f_{514}\rvert$ |
|---|---|---|---|---|
| A1 | 46.4 | 8.35 | 2.9 | 1.39 |
| A1 | 75.4 | 4.66 | 2.15 | 1.04 |
| A1 | 88.4 | 1.7 | 1.3 | 2.74 |
| A1 | 102.2 | 1.3 | 1.14 | 2.05 |
| A1 | 152.5 | 2.51 | 2.11 | 1.58 |
| A1 | 164 | 3.0 | 2.16 | 1.73 |
| A1 | 172.5 | 7.12 | 2.67 | 4.09 |
| A1 | 194.4 | 3.74 | 1.94 | 2.3 |
| A1 | 223.8 | 2.5 | 1.58 | 1.13 |
| A1 | 261.5 | 6.48 | 2.54 | 2.54 |

| Symmetry | f ($cm^{-1}$) | $I_{633}/I_{514}$ | $\lvert h_{633}\rvert/\lvert h_{514}\rvert$ |
|---|---|---|---|
| A2 | 125 | 2.5 | 1.58 |
| A2 | 184 | 2.8 | 1.67 |

**Table 2:** Enhancements factors ($|d_{633}|/|d_{514}|$ and $|f_{633}|/|f_{514}|$) for Raman tensor elements |d|, |f| and |h| as determined from $I_{633}/I_{514}$ for both $A_1$ and $A_2$ modes and |d|/|f| values for $A_1$ modes taken from angular resolved polarized Raman measurements at 633 nm and 514 nm listed in Table 1.

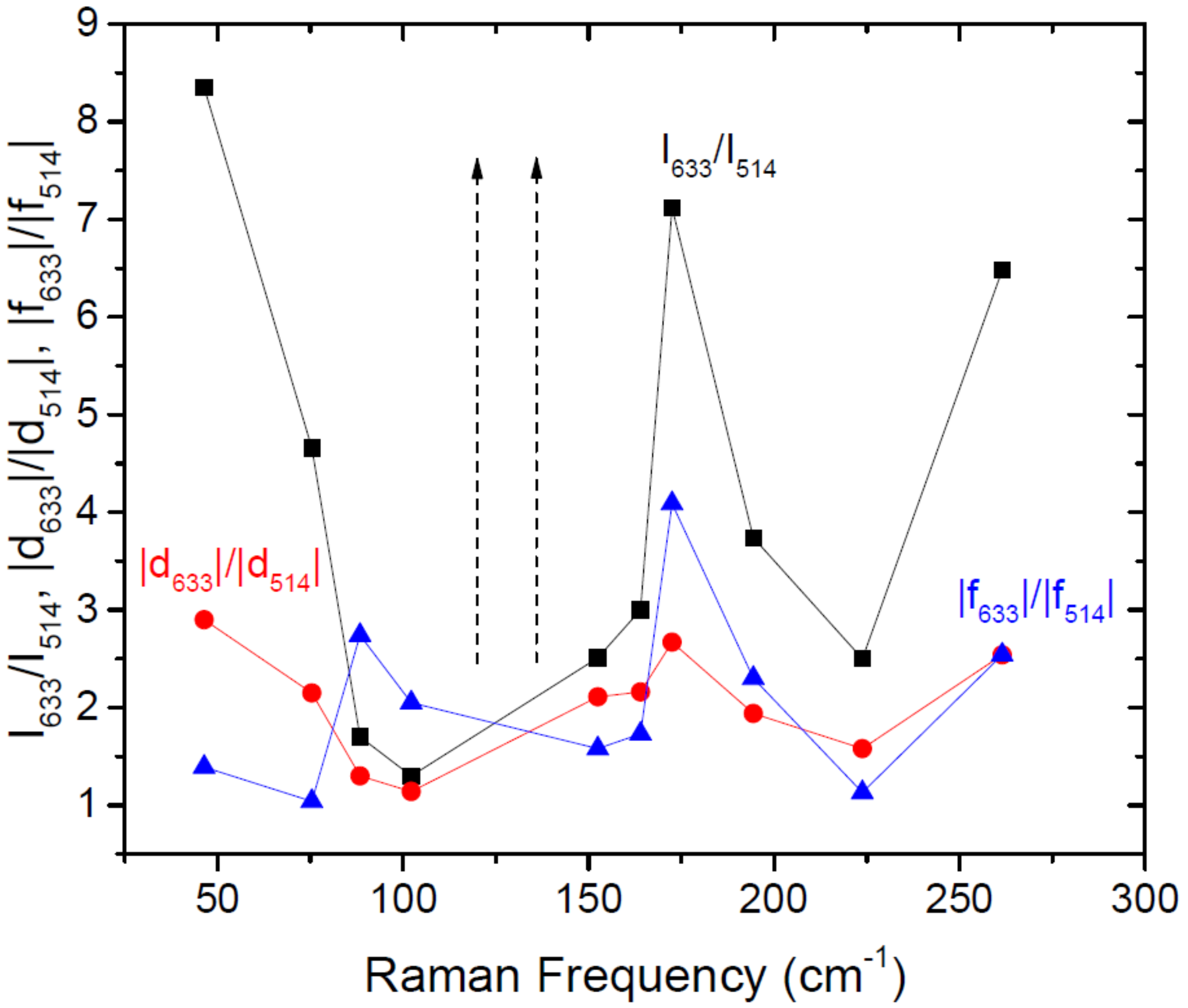

Figure 6. Plot of $I_{633}/I_{514}$ (black sqares), $|d_{633}|/|d_{514}|$ (red circles), and $|f_{633}|/|f_{514}|$ (blue triangles) as a function of phonon energy for $A_1$ modes. Vertical arrows show modes which are only seen for 633 nm excitation.

While all the $A_1$ phonon modes essentially have the same symmetries, the modes show a wide range of responses in both intensity and changes in the relative electron-phonon coupling parameters, |d| and |f|. Most modes show a substantial increase in Raman scattering efficiency in the red which range from factors of 2 to 8 (see Table 1 and 2 or Fig. 6), while two $A_1$ modes are not seen at all in the green, and two modes do not show any intensity enhancement. Such differing behavior from mode to mode is *not* what is usually seen in resonant Raman experiments. In most cases, all Raman modes are seen to increase in intensity substantially when a virtual intermediate *electronic* state is created and destroyed in the Raman scattering process because the energy denominator in the Raman tensor element tends towards zero for all modes (see denominator of equation 1). One concludes that the variation in the intensity and also in |d/|f| must result from differences in the matrix elements in the numerator. An incoming photon creates a real or virtual

electronic state which creates a phonon through electron-phonon coupling by scattering to an intermediate electronic state before the electron and hole recombine emitting a photon at a lower energy. To show the wide variation in the energy variation of the Raman tensor elements, we use equations 8-11 to extract $|d_{633}|/|d_{514}|$, $|f_{633}|/|f_{514}|$ and $|h_{633}|/|h_{514}|$. These values are recorded in Table 2. In fig. 6 we display $I_{633}/I_{514}$, $|d_{633}|/|d_{514}|$, and $|f_{633}|/|f_{514}|$ as a function of phonon energy for the $A_1$ phonon modes. One can clearly see strong increases in the Raman scattering efficiencies for red (633 nm) excitation in many of the Raman modes. However, the behavior of |d| and |f| are different. For example, maximum intensity enhancements are seen for the phonon frequencies at 46.4, 172.4 and 261.2 $cm^{-1}$. The 46.4 $cm^{-1}$ mode shows that the enhancement for the |d| tensor element is larger than for |f|. However, the 172.4 $cm^{-1}$ mode shows that the enhancement for the |f| tensor element is larger than for |d|, while the 261.2 $cm^{-1}$ phonon shows the same enhancement for |d| and for |f|. More interestingly, there is a peak enhancement in the |f| tensor element where the intensity enhancement is a minimum. It is the variation of the electron-phonon coupling which results in the creation and scattering of phonons which results in such a wide variation in intensities and angular polarization response moving from green to red excitation. A number of theoretical works have shown that the coupling of phonons to the chiral Fermions in these topological Weyl semimetals is quite complex.[29], [30], [32], [33] Substantial changes (even Fano resonances) have been observed as a function of temperature for both Raman scattering and also infrared absorption, and also most recently in a magnetic field.[34]–[38], [40] The results here shows that substantial changes can also occur in the electron-phonon coupling during the creation of virtual intermediate states in resonant Raman scattering.

**Conclusion**

A total of 19 Raman active modes, 13 $A_1$ and 6 $A_2$ from $NbIrTe_4$ have been detected through angle-resolved polarized micro-Raman scattering for 633 nm and 514 nm excitations. The frequency and symmetry of the detected modes is consistent with expectations from DFT calculations. The DFT calculations also provide the normal modes for each mode in which $A_1$ modes have out-of-plane motions while $A_2$ modes have in-plane motions. Utilizing the symmetry of the $A_1$ modes, we are able to explore the relative change of the scattering efficiency and |d|/|f| electron-phonon constants as a function of excitation energy for each phonon mode. We observe efficiency increases for the majority of modes for red excitation rather than green, but these increases vary substantially from mode to mode. For example, some modes are unexpectedly absent for green excitation while others are seen to be unchanged for both red and green excitation. Thus, some modes clearly show increases or decreases of the |d|/|f| electron-phonon coupling parameters, while others do not change. Through analysis of these results, we are able to extract the quantitative enhancements of the electron-phonon tensor elements |d|, |f| and |h| for 10 $A_1$ phonons and 2 $A_2$ phonons ranging from 5 meV to 32 meV (45 to 260 $cm^{-1}$). We conclude that these changes cannot be related to common excitations of the same intermediate electronic states, but rather result from the complex coupling of the phonons with the complex electronic states (possibly the Chiral Fermion states) which have been seen in a number of other systems.[34]–[38]

**Methods:**

***NbIrTe4 Sample Preparation:*** Single crystals of $NbIrTe_4$ were synthesized *via* the self-flux method. Nb powder (Alfa, 99.99%), Ir powder (Alfa, 99.95%), and Te shot (99.999%) were sealed in a fused silica ampoule under $10^{-6}$ Torr of vacuum. The reagents were mixed to yield a solution of 5 at.% $NbIrTe_4$ in Te. The samples were heated to 1000C at a rate of 200C/h, soaked at 1000C for 24h, and subsequently cooled to 500C at 2C/hr. Molten Te was centrifuged at 500C to isolate crystals. Crystals present as thin, silver flakes with a metallic luster. Dimensions of 1mm x 2mm x 0.1 mm can be achieved under these conditions.

***Sample Characterization:*** Lattice parameters of $NbIrTe_4$ were obtained at 300K using single crystal X-ray diffraction (SCXRD). Data was collected using a Bruker Kappa Apex II single-crystal diffractometer with Mo Ka radiation and a TRIUMPH monochromator.

***Angular-Dependent Raman Scattering Measurements:*** Fig. 1d illustrates a schematic diagram of our polarized micro-Raman experiment system. For excitation we have used a polarized 632.8 nm wavelength of a He/Ne laser and a polarized 514.5 nm wavelength of an Argon-ion laser. The laser beam is focused to a ~1.5 µm spot size onto the nanoflake by using a 100 X objective. The laser power at 633 and 514 nm measurements is limited to less than 200 µW on the sample. The incoming and backscattered laser lights are along the 'c' axis of the flake and the linear polarization of the incident beam is fixed during experiment. By rotating the measurement stage which carries $NbIrTe_4$ flake on a $Si/SiO_2$ substrate, it is possible to align the incident laser polarization along the 'a' or 'b' axes of the crystal, or any arbitrary angle in between while keeping the laser spot constant on the nanoflake. The backscattered light passes through a polarizer which is fixed to provide the maximum efficiency through a Dilor triple-spectrometer with two 1800 $mm^{-1}$ gratings used in the subtractive mode to remove the scattered laser light followed by a dispersive spectrograph with a 1800 $mm^{-1}$ grating. The signal is detected by a LN2 cooled CCD camera. A half-wave plate before the polarizer is allows us to select the polarization of the scattered light ($e_s$) to be either parallel ($e_i \parallel e_s$) or perpendicular ($e_i \perp e_s$) to the incident laser polarization ($e_i$).

***DFT Calculations of Phonon Modes:*** Our calculations are performed using density functional theory (DFT) as implemented in the Vienna ab initio simulation package (VASP) code.[50]–[52]. The Perdew-Burke-Ernzerhof (PBE) exchange-correlation functional and the projector-augmented-wave (PAW) approach are used. Throughout the work, the cutoff energy is set to be 550 eV for expanding the wave functions into plane-wave basis, and the number of k points was set to 4×4×4 for a 3×1×1 supercell. The real-space force constants of the supercells were calculated in the density-functional perturbation theory (DFPT)[53] and the phonon frequencies were calculated from the force constants using the PHONOPY code.[54] In our calculations, we adopt the experimental structural parameters (a=3.7903Å, b=12.5207Å, and c=13.1435Å).

**Acknowledgements:**

We acknowledge the financial support of NSF through grants DMR 1507844, DMR 1531373, DMR 1505549 and ECCS 1509706, and the NSF of China through grant 11674278. S.D.W. and B.R.O. acknowledge support from the University of California Santa Barbara Quantum Foundry, funded by the National Science Foundation (NSF DMR-1906325). Research reported here also made use of shared facilities of the UCSB MRSEC (NSF DMR-1720256). B.R.O. also acknowledges support from the California NanoSystems Institute through the Elings Fellowship program.

**Author Contributions:**

IAS assisted by SP was primarily responsible for taking the data, analyzing the results and preparation of the figures. BRO and SDW were responsible for preparation of the samples and XRD measurements. CL and FZ were responsible for the DFT calculations and analysis. GJ, HEJ and LMS conceived the experiments, and supervised the analysis. IAS, HEJ and LMS wrote the manuscript. All authors reviewed the manuscript.

**Correspondence** and request for materials should be addressed to L.M.S. (email: leigh.smith@uc.edu).

## A Raman Probe of Phonons and Electron-phonon Interactions in $NbIrTe_4$

*Iraj Abbasian Shojaei [a], Seyyedesadaf Pournia, Congcong Le [b,c], Brenden R. Ortiz [d,e],*
*Giriraj Jnawali [a], Fu-Chun Zhang [b], Stephen D. Wilson [d,e],*
*Howard E. Jackson [a], Leigh M. Smith* [a],*
[a] *Department of Physics, University of Cincinnati, Cincinnati, OH, USA*
[b] *Kavli Institute of Theoretical Sciences, University of Chinese Academy of Sciences, Beijing 100190, China*
[c] *Max Planck Institute for Chemical Physics of Solids, 01187 Dresden, Germany*
[d] *Materials Department, University of California Santa Barbara, Santa Barbara CA 93106*
[e] *California Nanosystems Institute, University of California Santa Barbara, Santa Barbara CA 93106*

*email: leigh.smith@uc.edu

# Supporting information

**S1: DFT Calculations of Phonon Modes**

The calculations are performed using density functional theory (DFT) as implemented in the Vienna ab initio simulation package (VASP) code.[1]–[3] The Perdew-Burke-Ernzerhof (PBE) exchange-correlation functional and the projector-augmented-wave (PAW) approach are used. Throughout the work, the cutoff energy is set to be 550 eV for expanding the wave functions into plane-wave basis, and the number of k points was set to 4 × 4 × 4 for a 3 × 1 × 1 supercell. The real-space force constants of the supercells were calculated in the density-functional perturbation theory (DFPT)[4] and the phonon frequencies were calculated from the force constants using the PHONOPY code.[5] In our calculations, we adopt the experimental structural parameters (a=3.7903A°, b=12.5207A°, and c=13.1435A°) which have been measured by XRD at the University of California at Santa Barbara by Professor Stephan Wilson's group.

Fig.1(a) in the main text shows crystal structure of $NbIrTe_4$, and all atoms occupy the same Wyckoff 2a{(0, y, z), (1/2, -y, z+1/2)} with corresponding site symmetry group $C_{1h}$. According to Characteristics of the table of $C_{1h}$, the vibrational direction of all $A_{1,2}$ and $B_{1,2}$ phonon modes are limited to "yz" plane or the "x" direction. Fig. S1(a) displays the phonon dispersion of $NbIrTe_4$, and no imaginary frequency is observed throughout the whole Brillouin zone. The density of states (DOS) of $NbIrTe_4$ is shown in Fig. S1(b), where the gray line is total DOS, and red, green and blue lines are the DOS of Nb, Te and Ir atoms, respectively. Range of frequency from 0 to 160 $cm^{-1}$, the DOS are mainly attributed to the Te atoms, and the peak around 187.202 $cm^{-1}$ is from contributions of Te and Ir atoms.

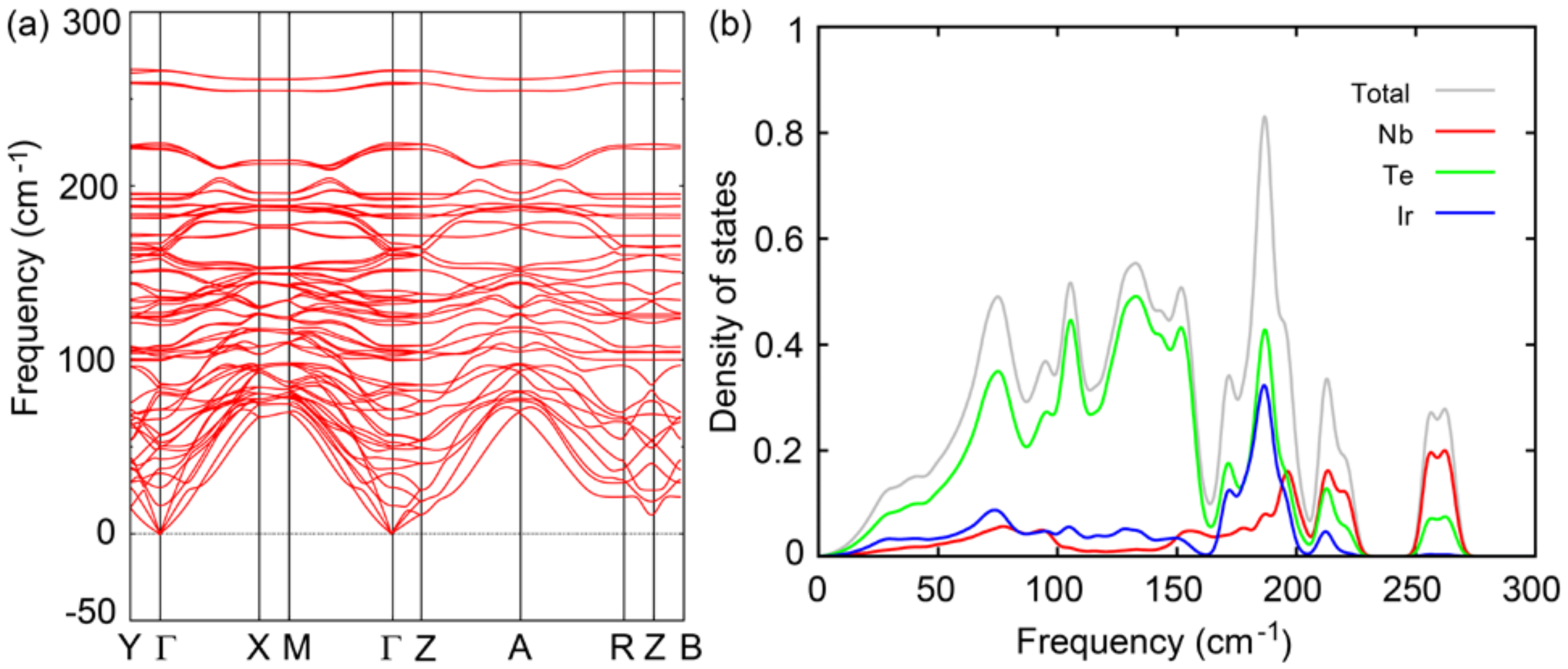

Figure S1: (a) and (b) are the phonon dispersion and DOS of $NbIrTe_4$. Gray is total DOS; Red, green and blue are the DOS of Nb, Te and Ir atoms, respectively.

## S2: Angular Dependence of Raman Scattering with |d|/|f|

Fig. S2 shows the angular behavior of three $A_1$ modes and one $A_2$ mode for both $e_i \parallel e_s$ and $e_i \perp e_s$ measurements. The left column shows a polar plot of the Raman intensities observed in $e_i \parallel e_s$ measurements and fit using equation (3), while the right column illustrates a polar plot of Raman intensities for $e_i \perp e_s$ measurements and fit to equation (4). The angular dependence of the $A_1$ modes for $e_i \parallel e_s$ measurements is substantially complex. The ratio of |d|/|f| and $\varphi_{df}$ have a significant role in determining the *shape* of the intensity as a function of θ (the angle between the laser polarization and the "x" axis of the lab coordinate system). As we see in the left column of the Fig. S2, modes with |d|/|f| > 1 (e.g. the 75.4 cm$^{-1}$ mode) have maxima for the laser polarization aligned with the "x" axis (θ = 0° and 180°), while modes with |d|/|f| < 1 (e.g. the 152.5 cm$^{-1}$ mode) have maxima for the laser polarized along the "y" axis (θ = 90° and 270°). If |d|/|f| = 1 and $\varphi_{df} = 0$, no variation of intensity with θ is observed (the shape is a circle). The 102.3 cm$^{-1}$ mode has |d|/|f| = 1 and $\varphi_{df} = 90°$ which exhibits four-lobed behavior with equal values along both the "x" and "y" directions. The angular dependence of all $A_1$ modes for the $e_i \perp e_s$ configuration, in contrast, shows simple four-lobed behavior with a maximum intensity aligned with 45° rotated "x" and "y" axes. As is predicted from theory, all $A_2$ modes display the expected four-lobe pattern as shown in the bottom row of Fig. S2, with the lobes aligned at 45° with respect to the "x" and "y" axes for the $e_i \parallel e_s$ measurements and aligned with the "x" and "y" axes for the $e_i \perp e_s$ configuration. From these observations, it is only possible to extract |d|/|f| for $A_1$ modes observed in the $e_i \parallel e_s$ configuration.

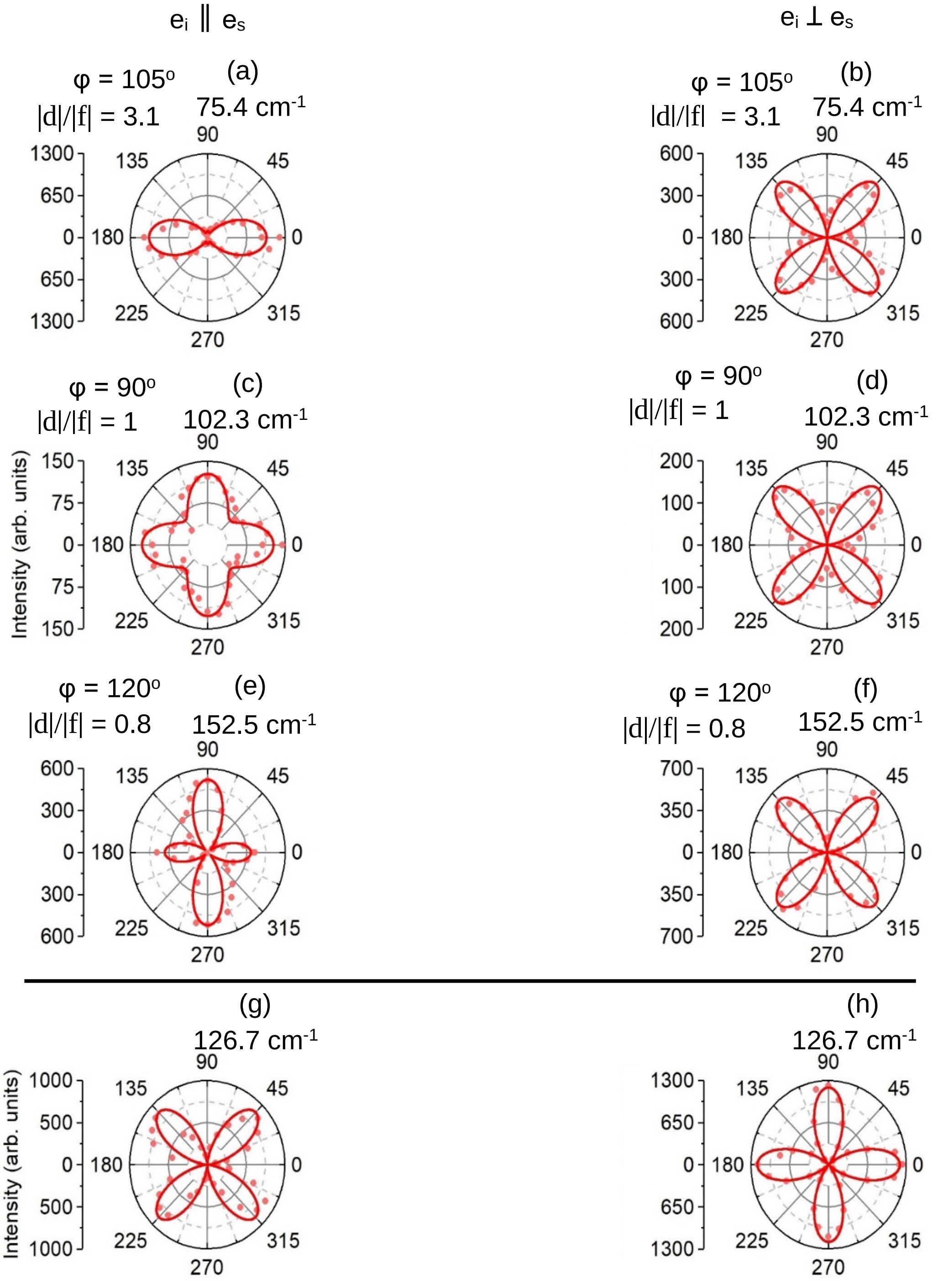

Figure S2: Intensity polar plot of four selected Raman modes of $NbIrTe_4$ as a function of rotation angle which have been measured with 633nm excitation. The first three rows belong to the $A_1$ irreducible representation and last row belongs to the $A_2$ irreducible representation. The left column represents a parallel configuration measurement where the solid red curves illustrate the theoretical fitting. The right column is a measurement for a perpendicular configuration with the solid red curves the theoretical fitting.

**S3: Angular Dependence of Raman Scattering with $\phi_{df}$.**

Fig. S3 shows three $A_1$ modes for both $e_i \parallel e_s$ and $e_i \perp e_s$ measurements. The left column shows a polar plot of Raman scattering intensities measured for $e_i \parallel e_s$ measurements which are fit to equation (3) while the right column illustrates a polar plot of Raman intensities measured for $e_i \perp e_s$ measurements and fit to equation (4). The phonon modes in this figure were chosen in order to illustrate the role of $\varphi_{df}$ for the angular behavior of $A_1$ modes for $e_i \parallel e_s$ measurements. Three $A_1$ modes are shown with similar $|d|/|f| \approx 1.5$ but different $\varphi_{df}$'s. One observes no change in the angular dependence (shape) of $e_i \perp e_s$ measurements (right column) by changing $\varphi_{df}$, but $e_i \parallel e_s$ measurements (left column) show a distinct variation with $\varphi_{df}$. All three modes of the $e_i \parallel e_s$ measurements show approximately two lobed behavior with a maximum intensity along the "x" axis ($\theta = 0^o$ and $180^o$) because $|d|/|f| > 1$. As $\varphi_{df}$ increases, at $\varphi_{df} = 130^o$, two small lobes appear which are aligned along the "y" axis ($\theta = 90^o$ and $270^o$). Generally, by increasing $\varphi_{df}$, two lobes perpendicular to the initial two lobes grow, and at $\varphi_{df} = 180^o$ it becomes a complete symmetric four lobed plot.

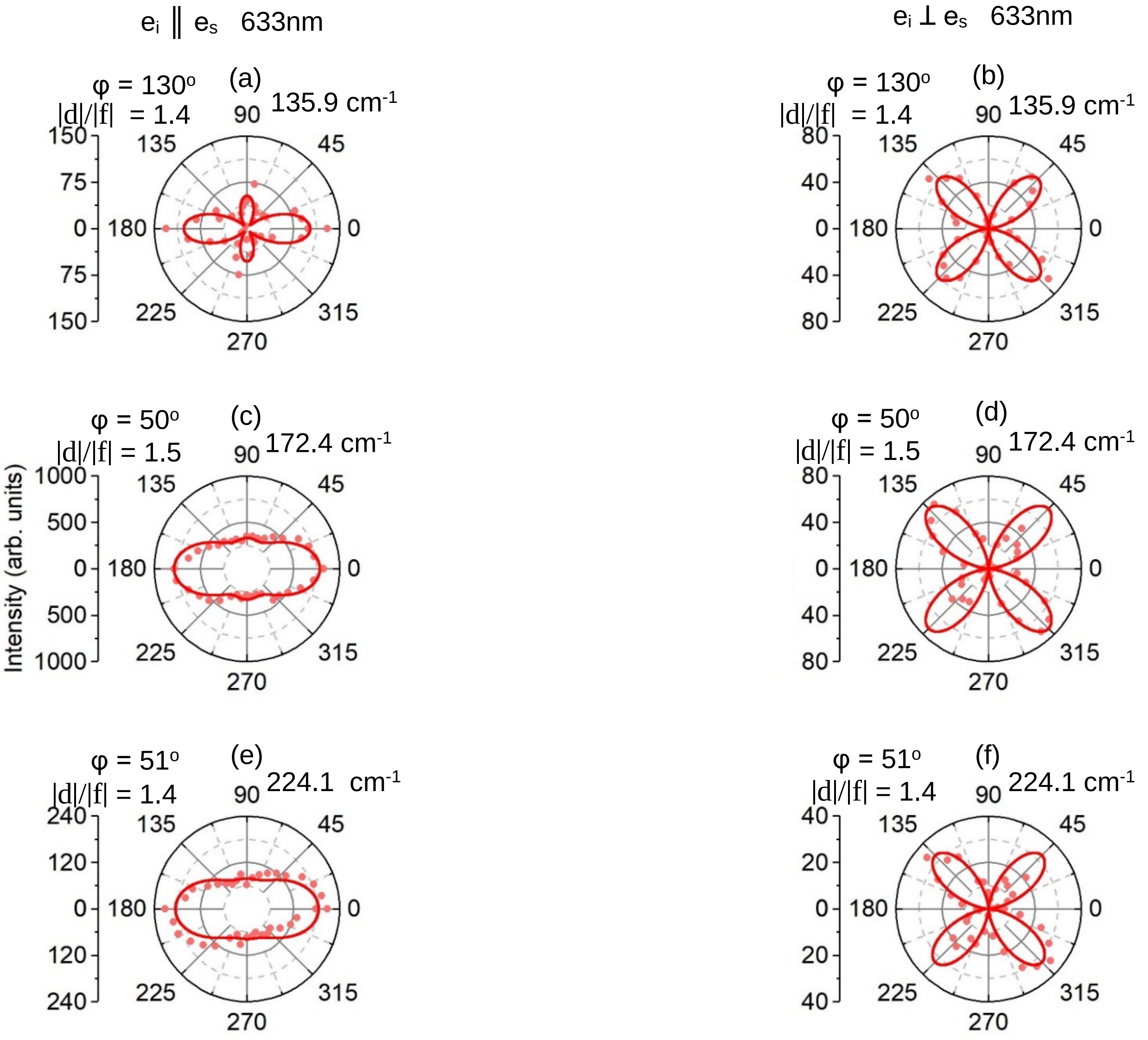

Figure S3: Intensity polar plot of three selected Raman modes of $NbIrTe_4$ with $A_1$ irreducible representation as a function of rotation angle with almost equal |d|/|f| but different $\varphi_{df}$ for fitting measurements using 633nm laser excitation. The left column shows parallel configuration measurements where the solid red curves illustrate theoretical fitting. Right column is measurement for perpendicular configuration with the red curves showing the theoretical fitting.

## S4: Spectral Calibration

We have used a Ne gas discharge source to calibrate the spectrometer. Fig. S4 shows the Ne spectrum measured by our spectrometer and CCD detector. We have detected 16 Ne emission lines for wavelengths longer than 514 nm to calibrate our Raman measurements when using a 514 nm excitation laser. Figure S5 shows two strong lines of a Ne gas discharge source for wavelengths longer than 633 nm to calibrate our Raman measurements when using a 633 nm excitation laser. The values of the peaks are shown in Table S1; we compare these values with values which are accepted $Ne^+$ lines in literature. The average difference between our measurements and literature values for the Ne lines is 0.018 nm or 0.67 $cm^{-1}$.

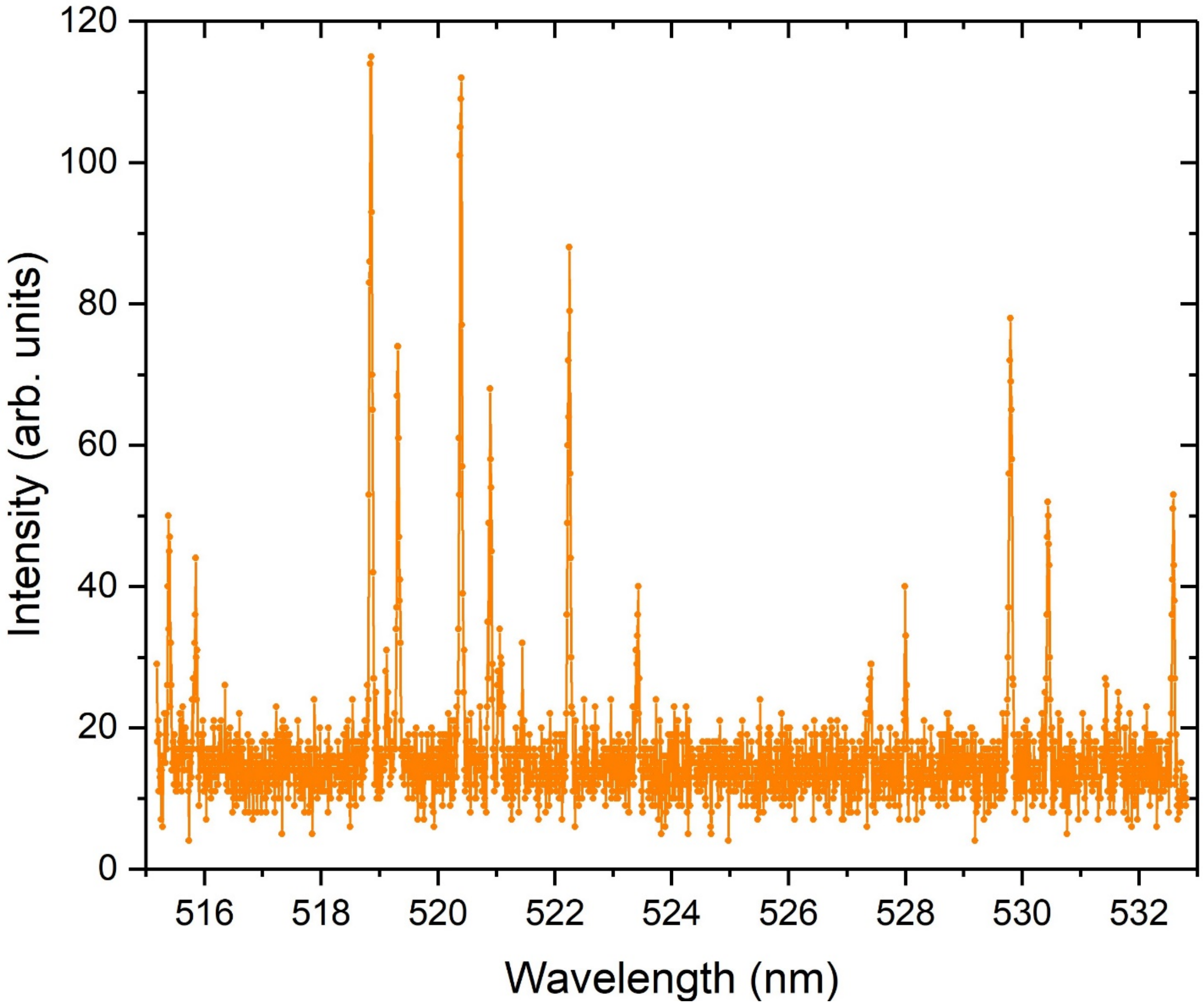

Figure S4: Atomic spectrum of the Ne gas discharge tube from 515 nm to 533 nm.

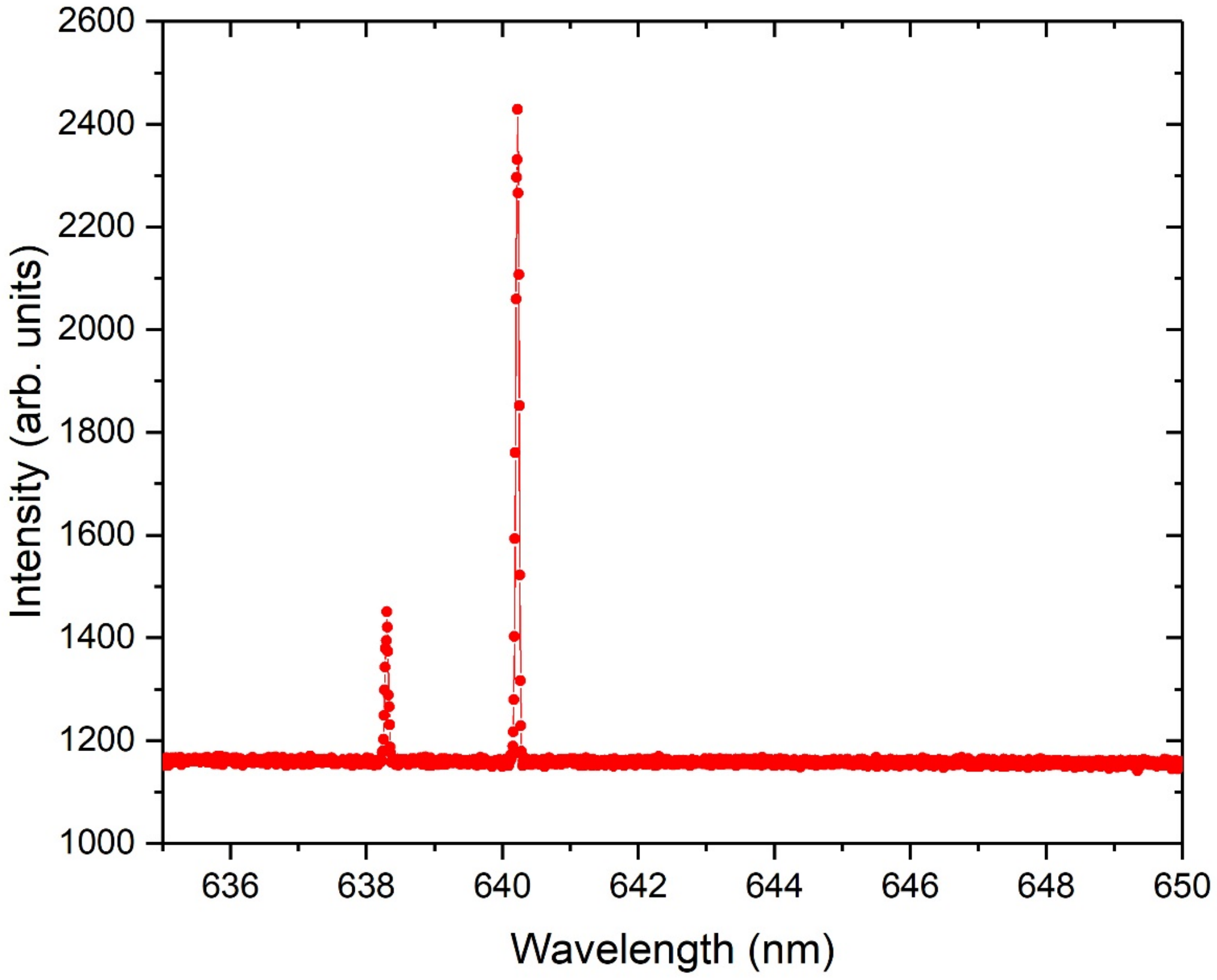

Figure S5: Atomic spectrum of the Ne gas discharge tube from 635 nm to 650 nm.

| No | Ne lines in our spectrometer (nm) | Ne lines in literature (nm) | Ne lines width in our spectrometer ($cm^{-1}$) | Measured line – literature line (nm) | Measured line – literature line ($cm^{-1}$) |
|---|---|---|---|---|---|
| 1 | 515.41382 | 515.44271 | 1.9 | -0.02889 | 1.09 |
| 2 | 515.85339 | 515.89018 | 1.6 | -0.03679 | 1.38 |
| 3 | 518.85602 | 518.86122 | 1.6 | -0.0052 | 0.19 |
| 4 | 519.12628 | 519.13223 | 1.5 | -0.00595 | 0.22 |
| 5 | 519.31799 | 519.31251 | 1.6 | 0.00548 | -0.2 |
| 6 | 520.4054 | 520.38962 | 1.3 | 0.01578 | -0.58 |
| 7 | 520.89166 | 520.88648 | 1.6 | 0.00518 | -0.19 |
| 8 | 521.05646 | 521.05672 | 0.5 | -0.00026 | 0.01 |
| 9 | 521.44659 | 521.43389 | 1.6 | 0.0127 | -0.47 |
| 10 | 522.25159 | 522.23517 | 1.5 | 0.01642 | -0.6 |
| 11 | 523.43439 | 523.40271 | 1.5 | 0.03168 | -1.16 |
| 12 | 527.42255 | 527.40393 | 0.9 | 0.01862 | -0.67 |
| 13 | 527.99377 | 528.00853 | 1.6 | -0.01476 | 0.53 |
| 14 | 529.80414 | 529.81891 | 1.4 | -0.01477 | 0.53 |
| 15 | 530.43951 | 530.47580 | 1.5 | -0.03629 | 1.29 |
| 16 | 532.59186 | 532.63960 | 1.3 | -0.04774 | 1.68 |
| 17 | 638.30054 | 638.29917 | 1.1 | 0.00137 | -0.03 |
| 18 | 640.22546 | 640.2248 | 1.1 | 0.00066 | -0.02 |

Table S1: The Ne atomic lines measured with our spectrometer and comparison with literature values

## S5: Raman Frequencies of $B_1$ and $B_2$ Phonon Modes

Tables S2 and S3 list all Raman active phonon modes of $NbIrTe_4$ with the $B_1$ and $B_2$ irreducible representation calculated by DFT. Table S2 shows the energy of 11 $B_1$ modes and Table S3 shows the energy of 23 $B_2$ modes.

| No | $cm^{-1}$ | No | $cm^{-1}$ | No | $cm^{-1}$ | No | $cm^{-1}$ |
|---|---|---|---|---|---|---|---|
| 1 | 50.188 | 4 | 100.348 | 7 | 127.055 | 10 | 181.463 |
| 2 | 57.038 | 5 | 104.987 | 8 | 160.848 | 11 | 183.437 |
| 3 | 65.903 | 6 | 124.203 | 9 | 163.368 | | |

Table S2: Calculated Raman active phonon energy with $B_1$ symmetry at Γ point by Density Function Theory.

| No | $cm^{-1}$ | No | $cm^{-1}$ | No | $cm^{-1}$ | No | $cm^{-1}$ |
|---|---|---|---|---|---|---|---|
| 1 | 34.719 | 7 | 105.950 | 13 | 151.020 | 19 | 195.510 |
| 2 | 37.384 | 8 | 108.060 | 14 | 158.093 | 20 | 221.688 |
| 3 | 48.814 | 9 | 122.498 | 15 | 164.224 | 21 | 223.131 |
| 4 | 80.062 | 10 | 132.031 | 16 | 171.826 | 22 | 258.347 |
| 5 | 83.788 | 11 | 133.861 | 17 | 188.621 | 23 | 266.455 |
| 6 | 86.271 | 12 | 136.206 | 18 | 192.993 | | |

Table S3: Calculated Raman active phonon energy with $B_2$ symmetry at Γ point by Density Function Theory.

## S6: Normal Modes and Angular Dependence for all observed Phonon modes

Fig. S6 shows the normal modes and angular dependence for all $A_1$ and $A_2$ phonon modes detected in our experiment. Results for both 514 and 633 nm excitation are shown for $A_1$ modes detected in the $e_i \parallel e_s$ configuration.

$A_1$

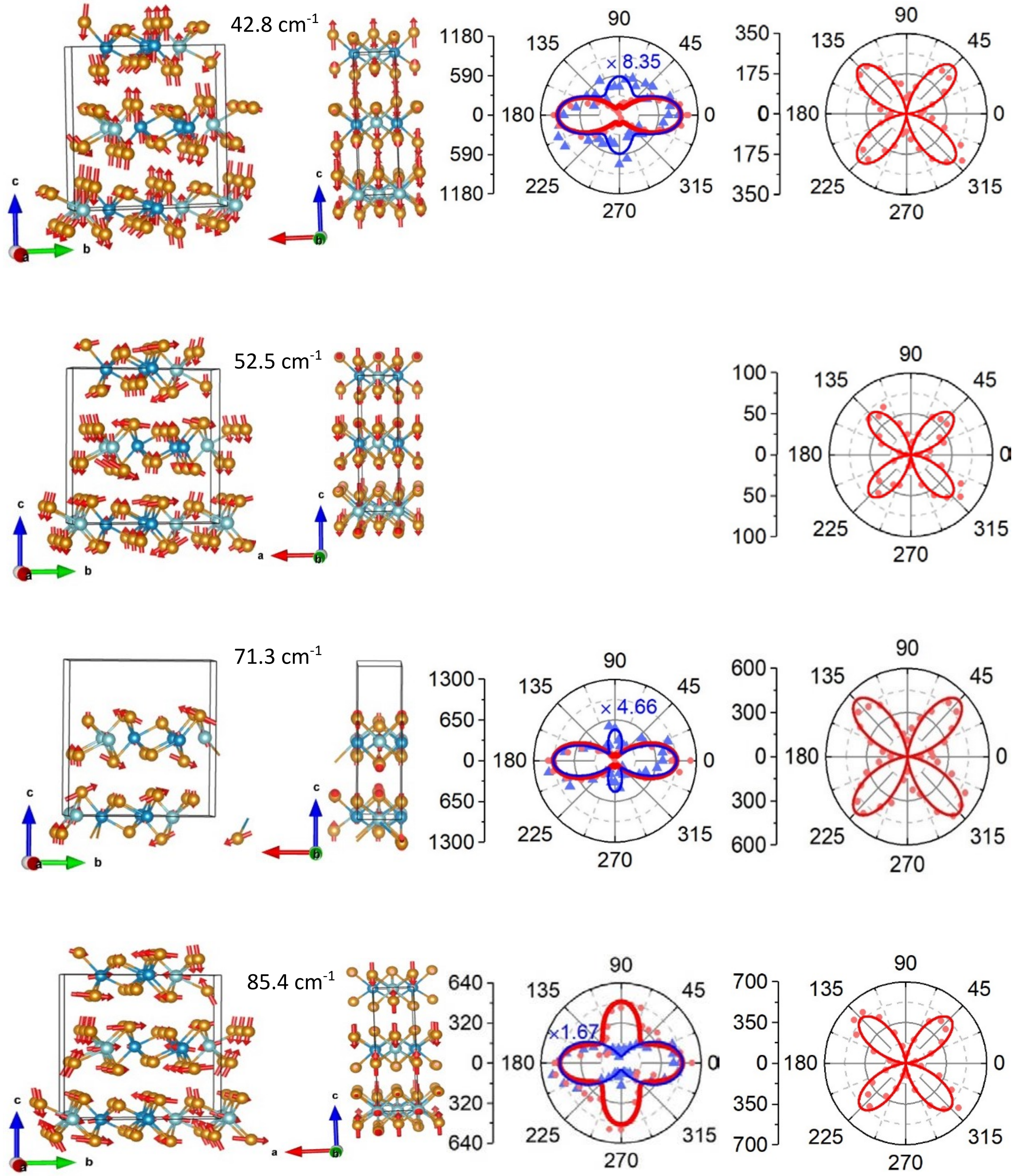

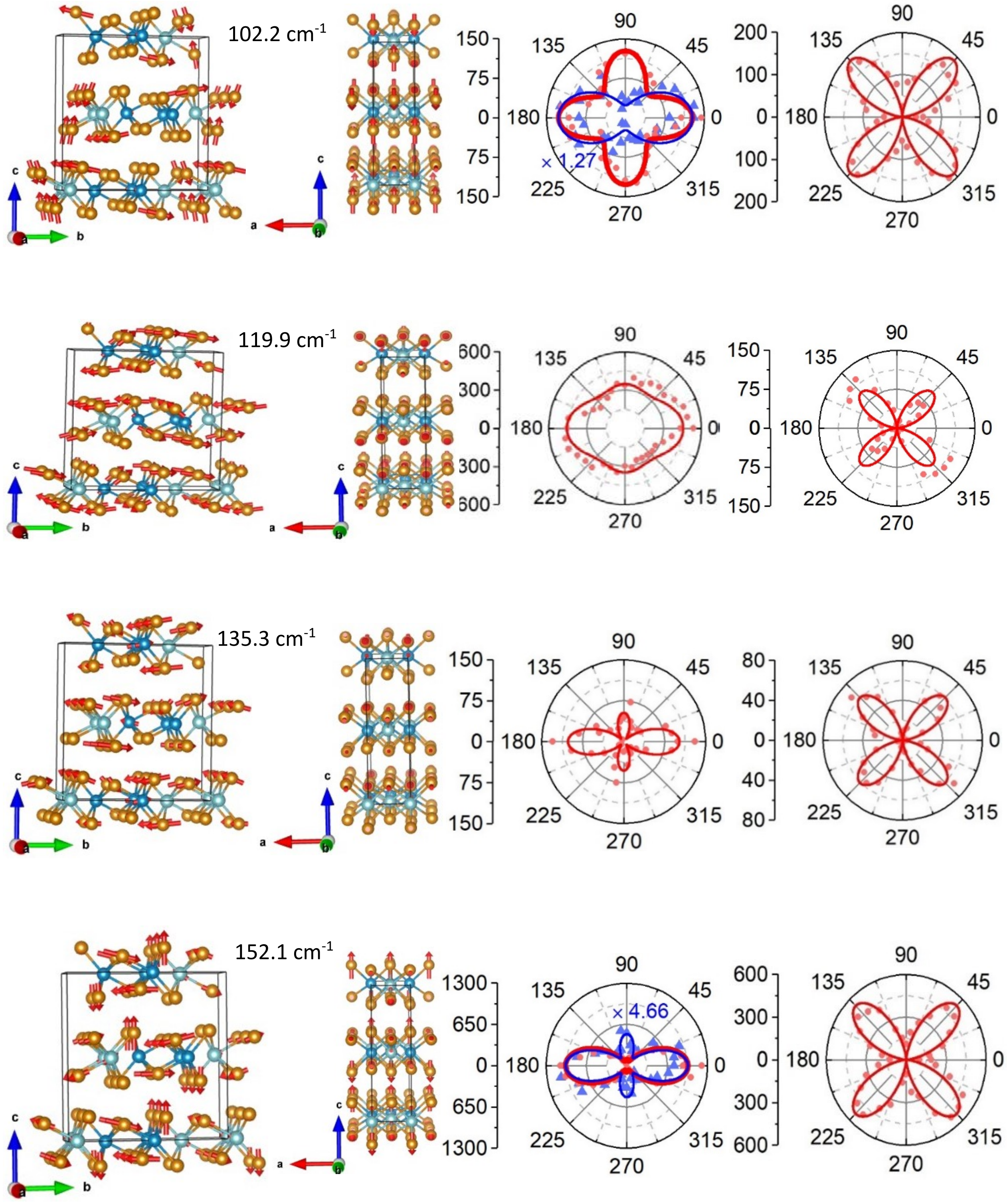

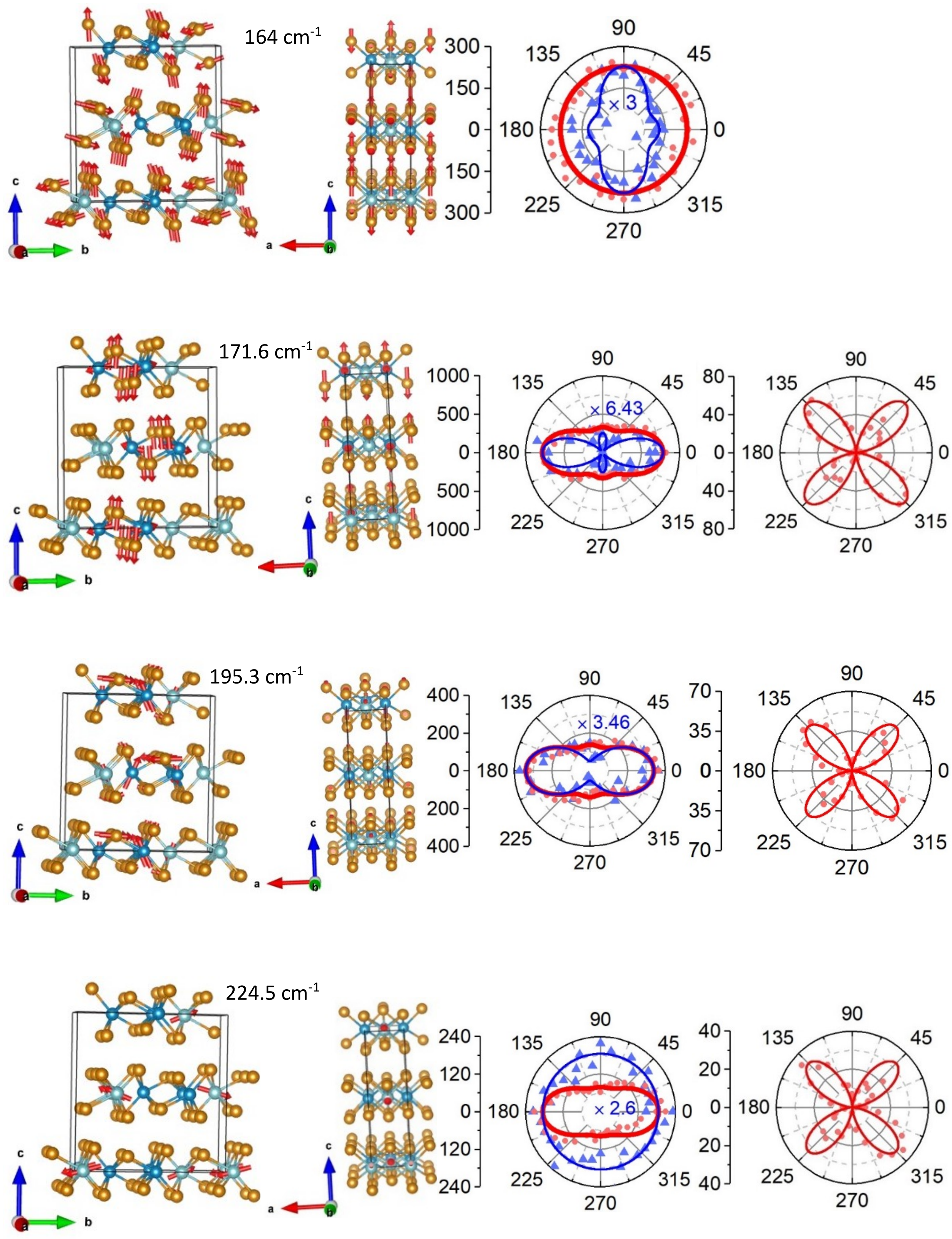

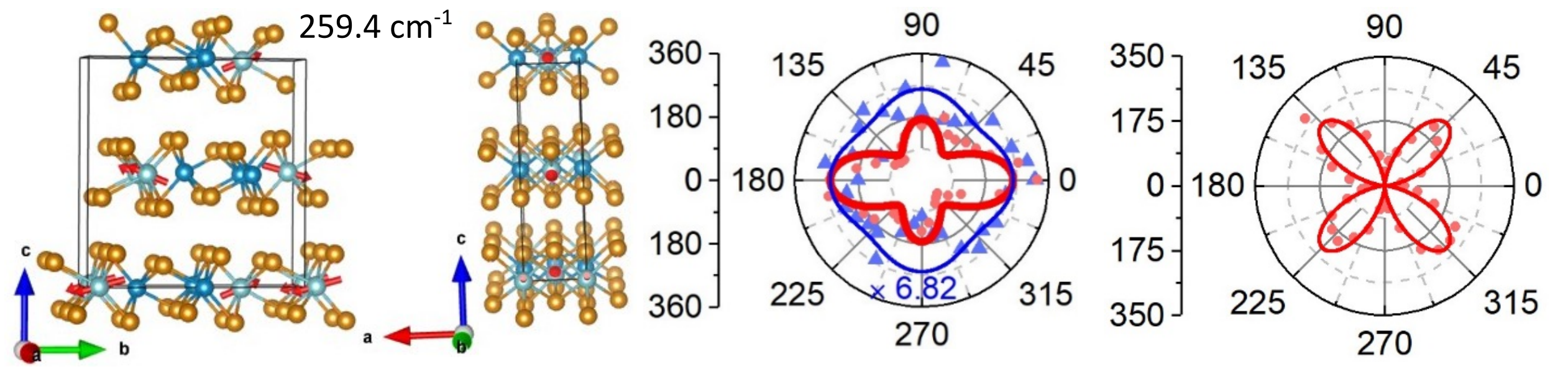

# A$_2$

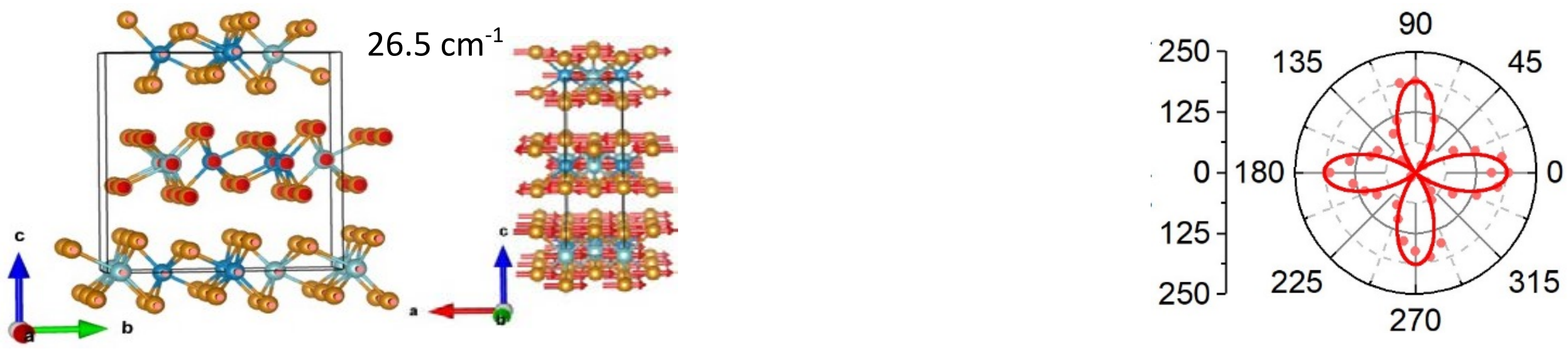

50.2 cm$^{-1}$

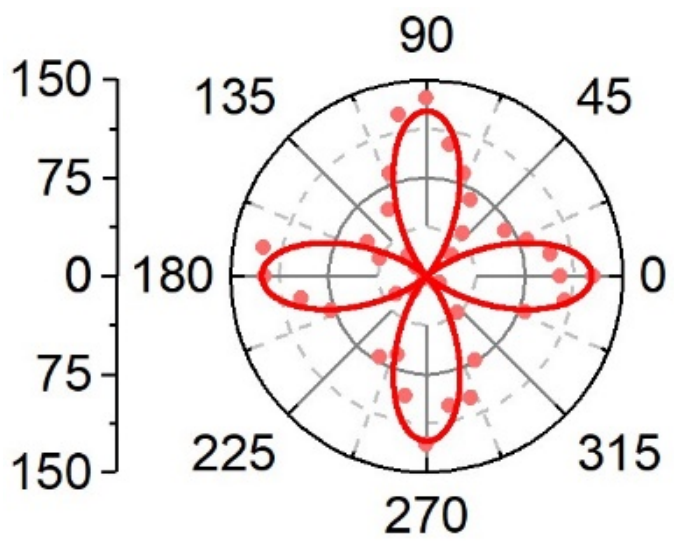

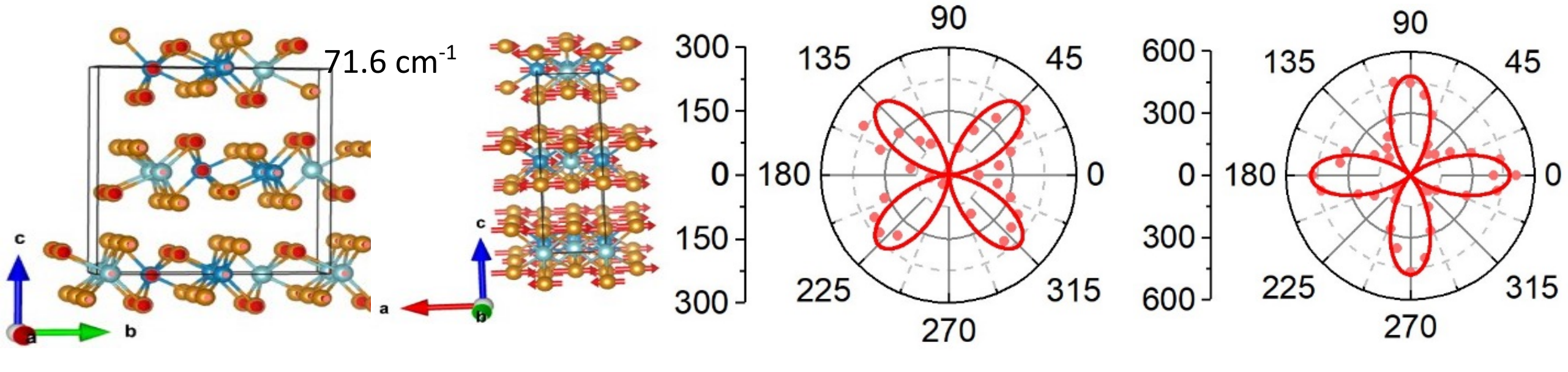

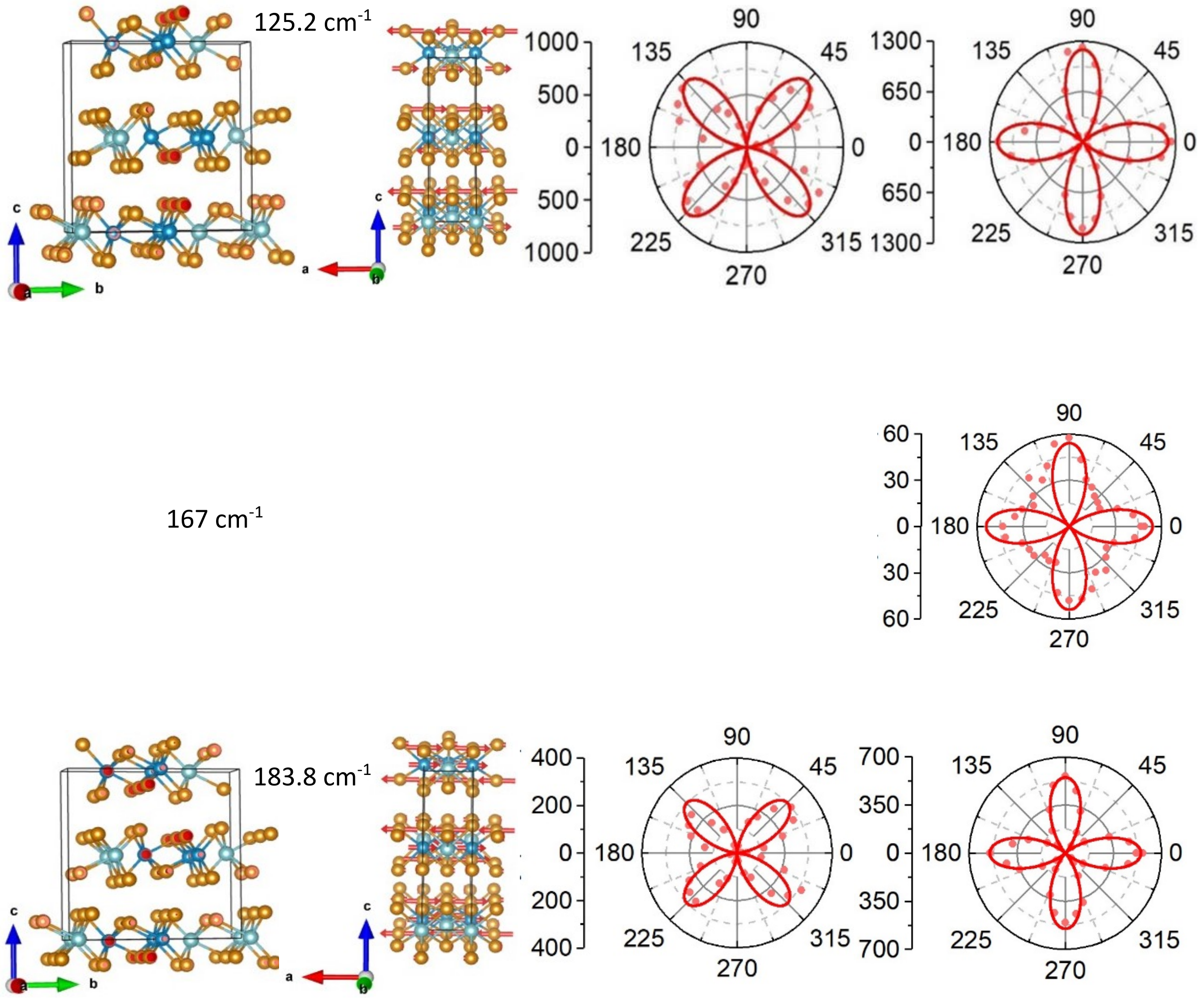

Figure S6: DFT calculation for normal modes of all $A_1$ and $A_2$ modes detected in our measurements in two perspective. Second and third columns are parallel and perpendicular measurements, respectively.

**S7: Determination of excitation wavelength dependence of Raman tensor elements**

We use the *shape* of the angular dependence of the Raman intensity at a particular laser excitation wavelength (either 514 or 633 nm) to determine the quantity |d|/|f| for each $A_1$ phonon mode. We also measure the change in the Raman intensity for each phonon mode at these two excitation wavelengths as $I_{633}/I_{514}$. Equation 3 in the main paper can be modified to read:

$$I_{A_1}^{\parallel} = |f|^2\left(\frac{|d|^2}{|f|^2}\cos^4(\theta) + \sin^4(\theta) + 2\frac{|d|}{|f|}\cos^2(\theta)\sin^2(\theta)\cos\left(\varphi_{df}\right)\right)$$

If the maximum occurs for θ = 0 (this occurs when d < f) then

$$I_{A_1}^{\parallel} = |f|^2\left(\frac{|d|^2}{|f|^2}\cos^4(0)\right)$$

and so one calculates the ratio of the intensities at 633 nm and 514 nm to be

$$\frac{I_{633}}{I_{514}} = \frac{|f_{633}|^2\,(|d_{633}|^2/|f_{633}|^2)}{|f_{514}|^2\,(|d_{514}|^2/|f_{514}|^2)}$$

Thus one obtains the final expression (removing the absolute value symbols for clarity):

$$\frac{|f_{633}|}{|f_{514}|} = \sqrt{\frac{I_{633}}{I_{514}}\frac{|d_{514}|/|f_{514}|}{|d_{633}|/|f_{633}|}}$$

If the maximum occurs for θ = π/2, then

$$I_{A_1}^{\parallel} = |f|^2\left(\sin^4(\pi/2)\right)$$

and so

$$\frac{|f_{633}|}{|f_{514}|} = \sqrt{\frac{I_{633}}{I_{514}}}$$

From this one can directly determine the change in the *d* Raman tensor element as:

$$\frac{|d_{633}|}{|d_{514}|} = \frac{|f_{633}|}{|f_{514}|}\frac{|d_{514}|/|f_{514}|}{|d_{633}|/|f_{633}|}$$